\documentclass[reprint,floatfix,amsmath,amssymb]{revtex4-2}
\usepackage[T1]{fontenc}
\usepackage[utf8]{inputenc}
\usepackage{bm,graphicx,booktabs}
\usepackage{microtype}
\usepackage[dvipsnames]{xcolor}
\usepackage{tikz}
\usepackage{pgfplots}
\pgfplotsset{compat=1.18}
\usepackage{hyperref}

\definecolor{cband1}{RGB}{40,116,166}
\definecolor{cband2}{RGB}{38,153,128}
\definecolor{cband3}{RGB}{221,132,38}

\hypersetup{hidelinks,
  pdftitle={Logit--Koopman coordinates for scale-resolved forecast-error growth in a three-level Lorenz-96 system},
  pdfauthor={Malaquias Pena}}
\newcommand{\sigmoid}{\operatorname{sigmoid}}
\newcommand{\logit}{\operatorname{logit}}
\newcommand{\dd}{\mathrm{d}}
\newcommand{\B}{\mathcal{B}}
\newcommand{\Sop}{\mathcal{S}}

\begin{document}
\title{Logit--Koopman coordinates for scale-resolved forecast-error growth
in a three-level Lorenz-96 system}
\author{Malaquias Pe\~{n}a}
\email{mpena@uconn.edu}
\affiliation{School of Civil and Environmental Engineering and
Eversource Energy Center, University of Connecticut,
Storrs, Connecticut 06269, USA}
\date{18 October 2026}

\begin{abstract}
Forecast-error growth in multiscale chaotic systems involves both
advancement toward saturation and evolution of the error's scale
distribution. We introduce Logit--Koopman coordinates to describe these
processes together. For error energies normalized
by fixed reference levels, logistic growth has exponential odds as exact
Koopman eigenfunctions and affine logits as growth clocks, with an
explicit inverse to error energy. Several band clocks define a
longitudinal coordinate for shared advancement and transverse
coordinates for departures from the reference band-growth relation.
An exact transformed error-energy budget resolves production,
conservative redistribution, interlevel exchange and dissipation into
contributions to clock speed and transverse motion; an equivalent
reference-law decomposition exposes the effect of ensemble
heterogeneity. In uncoupled and coupled three-level Lorenz-96 regimes,
200 forecasts per regime exhibit nearly affine band-mean logits
($R^2=0.965$--$0.993$), alongside distinct timing, residual and
launch-anomaly correlations. Recovered error fields reveal net
slow-to-fast interlevel error-energy exchange in the coupled control,
which removes slow-level error energy at a nearly steady relative
rate near $1.2$ per model time unit. Fast-only initial perturbations nevertheless
seed slow-level errors immediately; after this transient, their mean
approximately follows the control growth curve delayed by about 0.7
model time units. The
slower middle-band clock reflects the full physical balance and an
error deficit maintained at the dominant wavenumbers, rather than the
strongest transformed interlevel sink. The contribution is a
prescribed reference geometry in which growth timing, relative scale
evolution and signed physical exchange become separately measurable.
\end{abstract}
\maketitle

\begin{quotation}
How quickly a forecast of a chaotic system loses skill depends on how
fast its errors grow and on how that growth is shared among scales of
motion. A long-standing description treats the total error as a
logistic curve that rises exponentially and then saturates. The logit of
such a curve is an exact clock that advances at constant speed. With
several scale bands, these clocks define a geometry: motion along a
common direction measures how far an error-growth episode has advanced,
whereas motion across it measures how relative scale evolution departs
from the reference growth pattern.
Written in these coordinates, the error-energy budget identifies the
physical processes that change the clock speed and those that bend the
trajectory. Forecast experiments with a three-level Lorenz-96 model,
with and without coupling between its slow and fast levels, demonstrate
the construction and identify signed exchanges of forecast-error
energy separately from the influence of initial perturbations.
\end{quotation}

%% ======================================================================
\section{Introduction}\label{sec:introduction}
The predictability of chaotic flows is limited by sensitive dependence on
initial conditions,\cite{Lorenz1963} but in systems with many
interacting scales the limit is set by how error propagates among
them.\cite{Lorenz1969,LeithKraichnan1972,RotunnoSnyder2008,Palmer2014}
Small-scale errors saturate quickly and can contaminate progressively
larger scales, so predictability depends on the spectral distribution
of error as well as on its total
size.\cite{Lorenz1969,DurranGingrich2014,BoffettaMusacchio2017}
Studies of forecast error accordingly operate at two levels: scalar
growth laws fitted to aggregate error
curves,\cite{Lorenz1982,DalcherKalnay1987,Savijarvi1995,SimmonsHollingsworth2002}
and scale-resolved diagnostics of evolving error
spectra.\cite{SelzCraig2015,Zagar2017,Judt2018,Zhang2019}
In dynamical-systems terms, the scale dependence of error growth is
quantified by finite-size Lyapunov exponents, which measure growth rates
at prescribed error amplitudes.\cite{Aurell1997,Boffetta2002}

These two descriptions are rarely joined. A logistic curve fitted to the
total error summarizes an aggregate lifecycle---exponential growth
followed by saturation---but the same aggregate can accompany different
spectral histories. Conversely, error spectra retain the scale
distribution but provide no common measure of how far an error-growth
episode has progressed. The question addressed here is how a small set
of physically defined error observables can describe both growth timing
and relative scale evolution, and how departures from a reference growth
law can be attributed to identifiable physical processes.

Koopman operator theory provides a natural setting, because it
represents nonlinear dynamics through the linear evolution of
observables.\cite{Koopman1931,Mezic2005,Budisic2012,Mezic2013,Brunton2022}
Data-driven methods approximate this evolution in supplied or learned
dictionaries, including dynamic mode
decomposition,\cite{Schmid2010,Rowley2009,Tu2014} its extended
form\cite{Williams2015} and deep networks that learn linearizing
coordinates.\cite{Lusch2018} For the logistic equation the Koopman
linearization is explicit: the odds are an eigenfunction and the logit
advances at constant speed. Morrison and Kutz construct the equivalent
eigenfunction, the reciprocal odds, for $\dot x=x^2-x$, the logistic
equation with reversed rate, from its invariant
manifolds.\cite{MorrisonKutz2024} We use this scalar
result as a prescribed, physically normalized observable rather than a
learned one, and develop its multiscale consequences: the geometry of
several growth clocks, the information carried by their contrasts and
correlations, and the exact transformation of physical error-energy
budgets into clock speed and transverse motion.

A companion study\cite{Pena2026NPGReduction} establishes when logistic
growth arises as a reduction of spectral error dynamics. Within a
bilinear exchange model, a stationary reference spectrum and a common
conversion ratio permit an invariant spectral ray with logistic
amplitude, and an exact two-shell construction shows that the same
aggregate can accompany distinct spectral histories. Here we develop
coordinates that retain those histories when band observations are
available and apply them to observing-system simulation experiments
(OSSEs) with a three-level Lorenz-96
system,\cite{Lorenz1996,Thornes2017} a standard testbed for multiscale
predictability and data
assimilation.\cite{LorenzEmanuel1998,Orrell2003,Wilks2005,Arnold2013}
Two regimes are compared: one in which the three levels are decoupled,
so that only the slow ring is active, and one with active coupling to
two faster levels. In each, 200 deterministic forecasts are launched
from analyses of a local ensemble transform Kalman filter
(LETKF).\cite{Hunt2007}

The approach differs from each of the three lines of work above.
Scalar growth laws are fitted to an aggregate curve and supply no
coordinates for its scale composition. Spectral error budgets resolve
production and transfer by scale but provide no common measure of
progress toward saturation in which those terms can be compared across
scales at different growth stages. Learned Koopman coordinates linearize
dynamics but do not by themselves prescribe a physical normalization or
an error-energy budget. Here a prescribed observable, the logit of a reference-normalized
error energy, plays all three roles: it is a clock, it is invertible to
physical error energy, and the exact error-energy budget transforms into
it term by term.

The contributions are as follows.
(i)~We characterize Logit--Koopman coordinates: the logit is the unique
constant-speed coordinate of the logistic reference up to an affine
change of scale and origin, and any choice of clock coordinate is
equivalent to a hypothesis about the scale-dependent growth rate
(Sec.~\ref{sec:theory}).
(ii)~For several scale bands we introduce a longitudinal--transverse
decomposition that separates shared growth timing from relative
spectral evolution, together with three correlation diagnostics that
distinguish mean timing, detrended departures and launch-to-launch
alignment (Secs.~\ref{sec:theory} and~\ref{sec:methods}).
(iii)~We derive an exact transformed error-energy budget, including the
ensemble-heterogeneity correction for clocks of ensemble-mean errors,
and its explicit form for the three-level Lorenz-96 system, in which
interlevel error-energy exchange cancels pairwise and triad transfer
among Fourier bands is conservative (Secs.~\ref{sec:budgettheory}
and~\ref{sec:errorbudget}).
(iv)~We apply these tools to the uncoupled and coupled OSSE regimes and
quantify coordinate linearity, evaluate the budget with re-integrated
forecasts to determine the direction of interlevel error-energy exchange,
test fast-to-slow error influence with initial-condition partitions, and
attribute the clock speed and transverse motion to physical terms
(Sec.~\ref{sec:results}). The principal physical finding is that, in the
coupled regime, net error-energy exchange runs from slow to fast levels
and acts on the slow-level error as a nearly steady additional damping,
whereas fast-level initial errors seed the slow level, whose mean error
then approximately follows the control growth curve with a nearly
constant delay.

Section~\ref{sec:discussion} positions the coordinates within
Koopman-based analysis, relates the clocks to finite-size Lyapunov
exponents, interprets multiscale error propagation in clock
coordinates, outlines applications beyond the present experiments and
states their limitations. Appendix~\ref{sec:probabilityappendix} gives a probability
interpretation of the coordinates. Throughout, ``spectral'' refers to a
Fourier or modal partition of the error field unless the spectrum of an
operator is stated explicitly.

%% ======================================================================
\section{Logit--Koopman coordinates}\label{sec:theory}
\subsection{Exact logistic reference}
Let $E\ge0$ be the error energy of a specified physical projection,
$L>0$ a fixed reference energy, and $q=E/L$. The reference dynamics are
\begin{equation}
 \dot q=\lambda q(1-q),\qquad \lambda>0,\quad 0<q<1.
 \label{eq:logistic}
\end{equation}
Define the logit and the odds,
\begin{equation}
 \eta=\logit(q)=\log\frac{q}{1-q},\qquad z=e^\eta=\frac{q}{1-q}.
 \label{eq:coordinates}
\end{equation}
For the Koopman family $\mathcal K^h f=f\circ\Phi^h$ of
Eq.~\eqref{eq:logistic}, where $\Phi^h$ is the time-$h$ flow map,
\begin{equation}
 \mathcal K^h z=e^{\lambda h}z,\qquad
 \mathcal K^h\eta=\eta+\lambda h.
 \label{eq:koopman}
\end{equation}
The odds are therefore an eigenfunction of $\mathcal K^h$ with
eigenvalue $e^{\lambda h}$ (generator eigenvalue $\lambda$), and $\operatorname{span}\{1,\eta\}$ is an invariant subspace
on which the evolution is affine. We refer to $(q,z,\eta)$ and their
multiband extension as \emph{Logit--Koopman coordinates}. For chaotic
flows the scalar logistic equation serves as a reference; we do not
assume that the resulting finite dictionary is invariant under the
Koopman operator of the full system.

In the reference equation $L$ is a carrying capacity. In the experiments
below $L$ is a fixed climatological second moment computed from a long
truth trajectory, and individual errors can exceed it, in which case the
real logit is undefined. Each transformed diagnostic therefore specifies
its population and interval, including whether averaging over forecasts
precedes the transformation.

\subsection{Constant-speed clocks and scale-dependent growth rates}
The logistic reference fixes the clock coordinate analytically. If a
continuously differentiable observable $h(q)$ advances at a constant
nonzero speed $c$ along every trajectory of Eq.~\eqref{eq:logistic},
then
\begin{equation}
 h'(q)\lambda q(1-q)=c
 \quad\Longrightarrow\quad
 h(q)=\frac{c}{\lambda}\logit(q)+d .
 \label{eq:uniqueclock}
\end{equation}
The logit is thus the constant-speed coordinate of this growth law, up
to scale and origin; $c=\lambda$ and $h(1/2)=0$ select $h=\logit$. The
statement is a characterization within a specified scalar law, since any
one-dimensional flow with nonzero velocity admits a local time
coordinate. Its usefulness here is that the clock, its inverse and its
relation to error energy are all explicit:
\begin{equation}
 q=\sigmoid(\eta)=\frac1{1+e^{-\eta}},\qquad
 0<\frac{\dd q}{\dd\eta}=q(1-q)\leq\frac14 .
 \label{eq:decoder}
\end{equation}
Every finite coordinate value maps to an interior error fraction, and
coordinate perturbations are not amplified in the normalized error
fraction $q$.

The characterization extends to any hypothesized growth law. For a
trajectory with $0<q<1$ define the finite-amplitude growth rate
\begin{equation}
 \Lambda=\frac{\dd\ln q}{\dd t}.
 \label{eq:Lambda}
\end{equation}
For an error amplitude $\delta=(qL)^{1/2}$, $\Lambda=2\,\dd\ln\delta/\dd t$;
finite-size Lyapunov analysis measures growth at prescribed amplitudes
using threshold-passage or rescaling protocols.\cite{Aurell1997,Boffetta2002}
The instantaneous rate below connects to that framework but is not itself
a measured finite-size Lyapunov exponent. The logit speed is
\begin{equation}
 \dot\eta=\frac{\Lambda}{1-q},
 \label{eq:etaLambda}
\end{equation}
and, more generally, a monotone coordinate $h(q)$ advances at constant
speed $c$ if and only if $\Lambda=c/[q\,h'(q)]$. Choosing a clock
coordinate is therefore equivalent to prescribing the growth-rate
profile $\Lambda(q)$. Exponential growth, $\Lambda=\lambda$, makes
$\ln q$ affine in time; logistic growth, $\Lambda=\lambda(1-q)$, makes
the logit affine; Gompertz growth,\cite{Gompertz1825}
$\Lambda=-\lambda\ln q$, makes $-\ln(-\ln q)$ affine; and growth at
constant energy rate, $\Lambda\propto q^{-1}$, makes $q$ itself affine.
The last case corresponds to the linear-in-time error-energy growth
associated with error propagation through an inertial range, where the
finite-size Lyapunov exponent scales as
$\delta^{-2}$.\cite{Aurell1997,BoffettaMusacchio2017} The
Dalcher--Kalnay law\cite{DalcherKalnay1987}
$\dot q=(\alpha q+s)(1-q)$ gives $\dot\eta=\alpha+s/q$, so a source
$s>0$ raises the logit speed at small error. Equation~\eqref{eq:etaLambda}
turns the choice of clock into a statement about $\Lambda$;
Sec.~\ref{sec:distinctive} compares the linearity of the corresponding
coordinates for the OSSE mean errors.

The construction is therefore not specific to the logit. For any
monotone growth law $\dot q=q\Lambda(q)$, the constant-speed coordinate
$h$ exists, $e^{h}$ is a Koopman eigenfunction of that law, and the
geometry and budget transformation below apply with the denominator
$L_bq_b(1-q_b)$ replaced by $L_b/h'(q_b)$, which is $L_bq_b(-\ln q_b)$ for Gompertz growth. In that case,
$-\ln q$ is itself an eigenfunction with generator
eigenvalue $-\lambda$. We adopt the logistic reference a priori, not by
goodness of fit, for four reasons: it is the quadratic saturation law of
classical predictability studies;\cite{Lorenz1982} it is the source-free
form of the Dalcher--Kalnay law, whose source those authors attributed
to model deficiencies,\cite{DalcherKalnay1987} which motivates it for
perfect-model experiments such as these, although projection onto a
subset of variables, interlevel coupling and averaging can still produce
effective departures from it (Sec.~\ref{sec:reflaw}); it is the reduction
derived from bilinear spectral exchange in the companion
study;\cite{Pena2026NPGReduction} and its denominator
$L_bq_b(1-q_b)$ is symmetric under $q\leftrightarrow1-q$ and coincides
with the logistic mobility of the probabilistic reference in
Appendix~\ref{sec:probabilityappendix}.

\subsection{Several clocks and their aggregate}\label{sec:severalclocks}
In the spectral reduction of the companion study,\cite{Pena2026NPGReduction}
the decorrelated error spectrum takes the form $\Delta(k,t)=a(t)U(k)$
with stationary $U>0$. A bilinear exchange model
$\partial_t\Delta=\mathcal B[U-\Delta,\Delta]+r_\Delta$ preserves this
ray from common-amplitude initial data and yields
$\dot a=\lambda a(1-a)$ when $r_\Delta$ vanishes on the ray and
$\mathcal B[U,U](k)/U(k)=\lambda>0$ on its support. Internal spectral
interactions may be present in $\mathcal B$; the condition concerns only
their action on the preserved shape. (The companion uses a half-energy
convention; doubling both error and reference leaves $q$ and its logit
unchanged.)

For independent blocks that satisfy these assumptions separately,
\begin{equation}
 \eta_b(t)=a_b+\lambda_b t,\qquad
 \mathcal K^h e^{\eta_b}=e^{\lambda_bh}e^{\eta_b},
 \label{eq:product}
\end{equation}
so several logistic components have an exact vector representation.
Their reference-weighted aggregate $A=\sum_bL_bq_b/\sum_bL_b$ is in
general a mixture of logistic curves; a single common-amplitude ray is
the special case of equal $q_b$. The aggregate logit $\eta_A=\logit(A)$
measures overall advancement, and the contrasts $\eta_b-\eta_A$ measure
relative normalized error levels; they vanish when all band amplitudes
coincide. Pairwise contrasts $\eta_i-\eta_j$ are constant for equal
rates with different offsets, whereas band-minus-aggregate contrasts
need not be, and unequal rates make all contrasts evolve. Retaining band
observations makes these distinctions measurable.

\subsection{Shared timing and transverse geometry}\label{sec:geometry}
Write Eq.~\eqref{eq:product} as
$\boldsymbol\eta(t)=\boldsymbol a+\boldsymbol\lambda t$.
Over any nondegenerate set of lead times,
\begin{equation}
 \operatorname{Cov}_t(\boldsymbol\eta)
 =\operatorname{Var}_t(t)\,\boldsymbol\lambda\boldsymbol\lambda^{\mathsf T}.
 \label{eq:covariance}
\end{equation}
The temporal covariance has rank one, and every pairwise correlation
equals one for positive rates, although the reduced equations are
independent. Strong temporal correlation of band clocks thus expresses
shared timing and, by itself, is not evidence of interaction. Pooling
trajectories with different launch offsets adds covariance, so the
identity does not apply to an arbitrary collection of logistic
trajectories.

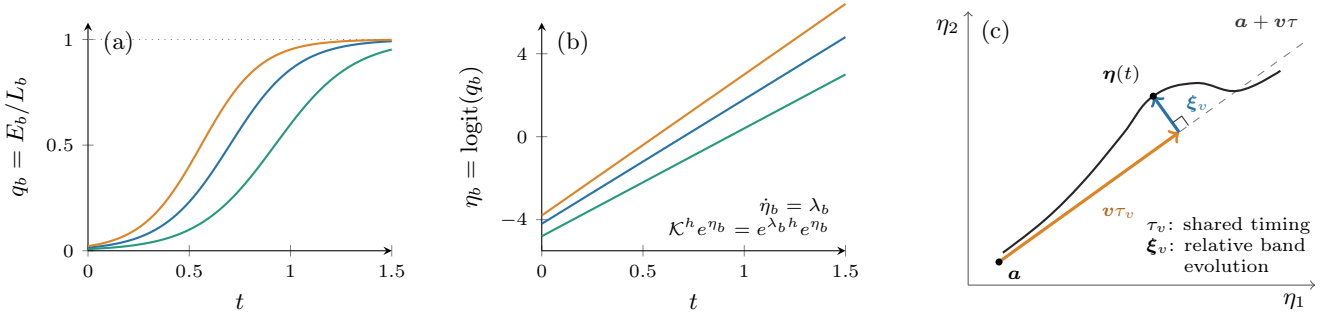
\begin{figure*}[t]
\centering
\begin{minipage}[b]{0.315\textwidth}\centering
\begin{tikzpicture}
\begin{axis}[width=5.6cm,height=4.6cm,xmin=0,xmax=1.5,ymin=0,ymax=1.08,
  axis lines=left,xlabel={$t$},ylabel={$q_b=E_b/L_b$},
  xtick={0,0.5,1,1.5},ytick={0,0.5,1},
  tick label style={font=\scriptsize},label style={font=\small},
  samples=120,domain=0:1.5,clip=false]
\addplot[cband1,thick] {1/(1+exp(-(-4.2+6.0*x)))};
\addplot[cband2,thick] {1/(1+exp(-(-4.8+5.2*x)))};
\addplot[cband3,thick] {1/(1+exp(-(-3.8+6.8*x)))};
\addplot[gray,dotted,domain=0:1.5] {1};
\node[anchor=north west,font=\small] at (rel axis cs:0.02,1.0) {(a)};
\end{axis}
\end{tikzpicture}
\end{minipage}\hfill
\begin{minipage}[b]{0.315\textwidth}\centering
\begin{tikzpicture}
\begin{axis}[width=5.6cm,height=4.6cm,xmin=0,xmax=1.5,ymin=-5.5,ymax=5.5,
  axis lines=left,xlabel={$t$},ylabel={$\eta_b=\logit(q_b)$},
  xtick={0,0.5,1,1.5},ytick={-4,0,4},
  tick label style={font=\scriptsize},label style={font=\small},
  samples=2,domain=0:1.5,clip=false]
\addplot[cband1,thick] {-4.2+6.0*x};
\addplot[cband2,thick] {-4.8+5.2*x};
\addplot[cband3,thick] {-3.8+6.8*x};
\node[anchor=north west,font=\small] at (rel axis cs:0.02,1.0) {(b)};
\node[anchor=south east,font=\scriptsize,align=right] at (rel axis cs:0.98,0.03)
  {$\dot\eta_b=\lambda_b$\\[1pt]
   $\mathcal K^h e^{\eta_b}=e^{\lambda_b h}e^{\eta_b}$};
\end{axis}
\end{tikzpicture}
\end{minipage}\hfill
\begin{minipage}[b]{0.33\textwidth}\centering
\begin{tikzpicture}[scale=1.02]
\draw[->,gray!70!black] (0,0) -- (4.5,0) node[below left,font=\small,black] {$\eta_1$};
\draw[->,gray!70!black] (0,0) -- (0,3.55) node[below left,font=\small,black] {$\eta_2$};
\draw[gray,dashed] (0.4,0.3) -- (4.35,3.144);
\node[font=\scriptsize,gray!40!black,anchor=south east] at (4.4,3.2) {$\boldsymbol a+\boldsymbol v\tau$};
\draw[black!85,thick] plot[smooth,tension=0.7] coordinates
  {(0.45,0.42) (1.0,0.85) (1.5,1.35) (2.0,1.95) (2.4,2.45) (2.95,2.62) (3.5,2.52) (4.05,2.78)};
\draw[->,very thick,cband3] (0.4,0.3) -- (2.7367,1.9824);
\draw[->,very thick,cband1] (2.7367,1.9824) -- (2.4,2.45);
\draw[gray!40!black] (2.8584,2.0700) -- (2.7708,2.1918) -- (2.6491,2.1041);
\fill (0.4,0.3) circle (1.3pt) node[below right,font=\scriptsize] {$\boldsymbol a$};
\fill (2.4,2.45) circle (1.3pt);
\node[font=\scriptsize,anchor=south east] at (2.35,2.5) {$\boldsymbol\eta(t)$};
\node[font=\scriptsize,cband3!80!black,anchor=north west] at (1.62,1.12) {$\boldsymbol v\tau_v$};
\node[font=\scriptsize,cband1,anchor=west] at (2.72,2.36) {$\boldsymbol\xi_v$};
\node[font=\small,anchor=north west] at (0.05,3.55) {(c)};
\node[font=\scriptsize,align=left,anchor=north west] at (2.2,1.0)
  {$\tau_v$: shared timing\\ $\boldsymbol\xi_v$: relative band\\ \phantom{$\boldsymbol\xi_v$: }evolution};
\end{tikzpicture}
\end{minipage}
\caption{Logit--Koopman coordinates for three scale bands (schematic).
(a) Normalized error energies $q_b=E_b/L_b$ that follow independent
logistic laws with different rates and offsets. (b) Their logits advance
at constant speeds $\lambda_b$; the odds $e^{\eta_b}$ are Koopman
eigenfunctions. (c) A band-logit trajectory $\boldsymbol\eta(t)$ (black)
relative to a reference line $\boldsymbol a+\boldsymbol v\tau$ (dashed),
shown in a two-band projection. The longitudinal coordinate $\tau_v$
measures advancement along the line, in units of time, and the
transverse vector $\boldsymbol\xi_v$ measures departure from the
reference band-growth relation [Eq.~\eqref{eq:geometry}]. For exact
independent logistic bands the trajectory lies on the line and
$\tau_v=t$.}
\label{fig:schematic}
\end{figure*}

For a fixed nonzero reference-rate vector $\boldsymbol v$, set
$\boldsymbol u=\boldsymbol v/\|\boldsymbol v\|$ and
$P_v=I-\boldsymbol u\boldsymbol u^{\mathsf T}$, and define
\begin{equation}
 \tau_v=\frac{\boldsymbol v^{\mathsf T}(\boldsymbol\eta-\boldsymbol a)}
                   {\|\boldsymbol v\|^2},\qquad
 \boldsymbol\xi_v=P_v(\boldsymbol\eta-\boldsymbol a),
 \label{eq:geometry}
\end{equation}
so that $\boldsymbol\eta=\boldsymbol a+\boldsymbol v\tau_v+\boldsymbol\xi_v$
(Fig.~\ref{fig:schematic}). In the exact benchmark,
$\boldsymbol v=\boldsymbol\lambda$ gives $\tau_v=t$ and
$\boldsymbol\xi_v=0$. The longitudinal coordinate $\tau_v$ measures
advancement along the reference line in units of time; the transverse
coordinates $\boldsymbol\xi_v$ measure departures from the reference
band-growth relation. A contribution parallel to $\boldsymbol v$ changes
the clock without bending the path. The aggregate logit $\eta_A$ is a
nonlinear amplitude summary whereas $\tau_v$ is a linear projection, and
a reference line with unequal rates has evolving band contrasts even
when $\boldsymbol\xi_v=0$. Covariance eigenvectors of these coordinates
summarize variability; they are not thereby physical normal modes or
Koopman eigenfunctions.

\begin{table*}[t]
\caption{Observable choices for the scalar logistic reference,
Eq.~\eqref{eq:logistic}. Generic dictionaries and learned coordinates
can contain the same functions; the comparison identifies the structure
that the present choice supplies explicitly. All statements assume
$0<q<1$ and a fixed reference $L$.}
\label{tab:coordinatechoices}
\small
\begin{tabular}{p{0.17\textwidth}p{0.22\textwidth}p{0.53\textwidth}}
\toprule
Representation & Reference evolution & Structure available for predictability analysis\\
\midrule
Normalized error $q$ & $\dot q=\lambda q(1-q)$ & Direct error fraction; nonlinear growth and saturation remain in the tendency.\\[3pt]
Odds $z=q/(1-q)$ & $\dot z=\lambda z$ & Exact Koopman eigenfunction with exponential propagation; inverse $q=z/(1+z)$ for $z>0$.\\[3pt]
Logit $\eta=\ln z$ & $\dot\eta=\lambda$ & Constant-speed clock, additive departures, and an inverse that maps every finite coordinate to $0<q<1$ with $\dd q/\dd\eta\le1/4$.\\[3pt]
Supplied or learned features & Dictionary-dependent approximation & Greater representational flexibility; physical meaning, invariance and reconstruction depend on the features chosen or learned.\\
\bottomrule
\end{tabular}
\end{table*}

\subsection{Transformed error-energy budget}\label{sec:budgettheory}
Let the band error energies obey an exact decomposition
$\dot E_b=\sum_m S_b^{(m)}$ into physically defined terms (production,
exchange, dissipation). On $0<q_b<1$ the chain rule gives
\begin{equation}
 \dot\eta_b=\sum_m\frac{S_b^{(m)}}{D_b},\qquad D_b=L_bq_b(1-q_b),
 \label{eq:logitbudget}
\end{equation}
so each term contributes additively to the logit speed. Any smooth
coordinate admits a chain-rule budget; the logit supplies the explicit
logistic clock and inverse in which to interpret these contributions.
Relative to the
logistic reference with rates $r_b$, write equivalently
\begin{equation}
 \dot E_b=r_b E_b(1-E_b/L_b)+C_b+R_b,
 \label{eq:budget}
\end{equation}
where $C_b$ is a physically defined exchange and $R_b$ collects the
remaining tendency; Eq.~\eqref{eq:budget} is exact once $C_b$ and $r_b$
are specified, and setting $R_b=0$ would be a closure. Then
\begin{equation}
 \dot\eta_b=r_b+\frac{C_b+R_b}{D_b},
 \label{eq:chain}
\end{equation}
and, with $D=\operatorname{diag}(D_b)$ and the projection of
Eq.~\eqref{eq:geometry} taken with $\boldsymbol v=\boldsymbol r$,
\begin{equation}
 \dot{\boldsymbol\xi}_r=P_rD^{-1}(\boldsymbol C+\boldsymbol R),
 \qquad
 \dot\tau_r=1+
 \frac{\boldsymbol r^{\mathsf T}D^{-1}(\boldsymbol C+\boldsymbol R)}
      {\|\boldsymbol r\|^2}.
 \label{eq:projectedbudget}
\end{equation}
The transformed budget separates physical contributions that change the
clock speed from those that bend the band-growth relation. Exchange that
conserves energy across a complete partition does not cancel among the
logit tendencies, because the denominators $D_b$ differ; production
remainders can also generate transverse motion.

For the logit of an ensemble-mean error, averaging must precede the
transformation. With common $r_b$ and $L_b$ across forecasts $j$,
\begin{equation}
 \frac{\dd}{\dd t}\logit(\bar q_b)=r_b+
 \frac{\overline C_b+\overline R_b-r_bL_b\operatorname{Var}_j(q_b)}
 {L_b\bar q_b(1-\bar q_b)},
 \label{eq:meanbudget}
\end{equation}
which follows from
$\overline{q_b(1-q_b)}=\bar q_b(1-\bar q_b)-\operatorname{Var}_j(q_b)$.
The variance term is nonpositive for positive reference rates and
quantifies a heterogeneity correction relative to the chosen logistic
reference. The remainder $R_b$ also depends on that reference, so this
correction is not a separately identified dissipative process. This
extends the companion study's emphasis on heterogeneous
amplitudes.\cite{Pena2026NPGReduction} Equivalently and without
reference to $r_b$, $\dd\logit(\bar q_b)/\dd t=\sum_m\overline{S}{}_b^{(m)}
/[L_b\bar q_b(1-\bar q_b)]$. Equations~\eqref{eq:logitbudget}--\eqref{eq:meanbudget}
give two equivalent decompositions. The complete physical budget
$\sum_m\overline S{}_b^{(m)}$ already contains all tendencies; the
variance correction belongs to the logistic-reference decomposition and
must not be added a second time to the physical budget. Section~\ref{sec:errorbudget} gives
these terms for the Lorenz-96 system.

%% ======================================================================
\section{Three-level Lorenz-96 OSSEs}\label{sec:methods}
\subsection{Model and regimes}
We use a three-level Lorenz-96 hierarchy\cite{Lorenz1996,Thornes2017}
with $K=36$ slow variables $X_k$, $J=10$ intermediate variables per slow
variable and $I=8$ fast variables per intermediate variable. With
$k(m)=\lfloor m/J\rfloor$ and $m(n)=\lfloor n/I\rfloor$,
\begin{equation}
 \begin{aligned}
 \dot X_k={}&X_{k-1}(X_{k+1}-X_{k-2})-X_k+F
              -\alpha\sum_{j=0}^{J-1}Y_{kJ+j},\\
 \dot Y_m={}&-c_xb_xY_{m+1}(Y_{m+2}-Y_{m-1})-c_xY_m\\
            &+\alpha X_{k(m)}-\beta\sum_{\ell=0}^{I-1}Z_{mI+\ell},\\
 \dot Z_n={}&-c_yb_yZ_{n+1}(Z_{n+2}-Z_{n-1})-c_yZ_n
              +\beta Y_{m(n)},
 \end{aligned}
 \label{eq:osse}
\end{equation}
with $\alpha=h_xc_x/b_x$, $\beta=h_yc_y/b_y$, $F=18$ and
$c_x=c_y=b_x=b_y=4$. Advection is periodic over the complete rings of
lengths $K$, $KJ$ and $KJI$. The parameters $c$ and $b$ separate the
levels in time scale and amplitude: $Y$ and $Z$ are damped four times
faster than $X$, and in the coupled regime their climatological standard
deviations are about 7 and 56 times smaller than that of $X$. In the
coupled regime (Run C) $h_x=h_y=1$ and all levels are active; in the
interlevel-uncoupled regime (Run U) $h_x=h_y=0$, the $Y$ and $Z$ levels
decay to zero, and only the $X$ ring is active. Its Fourier modes still
interact through the quadratic advection.

For $H=\tfrac12(\|X\|^2+\|Y\|^2+\|Z\|^2)$, advection conserves energy
on each ring and the coupling terms cancel pairwise, giving
$\dot H=F\sum_kX_k-\|X\|^2-c_x\|Y\|^2-c_y\|Z\|^2$. The signed
state-energy transfers $T_{XY}=\alpha\sum_mX_{k(m)}Y_m$ and
$T_{YZ}=\beta\sum_nY_{m(n)}Z_n$ are positive for $X\to Y$ and $Y\to Z$, respectively.
In Run C, evaluated on 2001 truth states spanning 200 model time units
(MTU), their means are 927.5 and 139.3 energy units per MTU, and both are
positive at every state; in Run U both vanish identically.

\subsection{Assimilation and forecasts}
Each regime is a perfect-model OSSE: truth and forecasts use the same
equations and parameters. Integration uses fourth-order Runge--Kutta
with step 0.001~MTU. A truth (nature) run of 200~MTU follows 100~MTU of
spin-up. The first 100~MTU are assimilated with an LETKF\cite{Hunt2007}
(40 members, Gaspari--Cohn localization with a half-width of four
slow-grid spacings) using observations every
0.05~MTU, and 200 deterministic forecasts of 6~MTU are launched from
analysis means between 20.00 and 99.95~MTU, approximately every
0.4~MTU (Table~\ref{tab:config}). Forecast errors are verified against
the truth every 0.01~MTU, giving 601 leads. Because launches share one
truth trajectory and their 6-MTU forecasts overlap, leads and launches
are not independent realizations. The regimes differ in climatology,
reference energies, observation-error magnitudes and initial errors, so
their comparison contrasts physically distinct regimes rather than an
intervention that changes only the coupling.

\begin{table}[t]
\caption{Configuration of the two OSSE regimes. $\sigma^2$ is the
climatological variance of a variable at the indicated level, and
$\bar q(0)$ the launch-mean normalized error at lead zero. $Y$
observations are assimilated in both runs but carry no information in
Run U, where $Y\equiv0$. $^\ast$The Run U multiplicative inflation
is inferred from the retained implementation; it was not recorded in
the run log.}
\label{tab:config}
\footnotesize
\begin{tabular}{@{}lll@{}}
\toprule
 & Run U & Run C\\
\midrule
Coupling $(h_x,h_y)$ & $(0,0)$ & $(1,1)$\\
Active variables & 36 ($X$) & 3276 ($X,Y,Z$)\\
Observed every 0.05 MTU & $X$ & $X$, $Y$\\
Obs.\ error variance & $0.01\sigma_X^2$ & $0.01\sigma_X^2$; $0.09\sigma_Y^2$\\
Inflation, mult.; add. & 1.15$^\ast$; $0.05\sigma_X^2$ & 1.15; $0.02\sigma^2$\\
$\sigma_X$; analysis RMSE of $X$ & 6.86; 0.55 & 4.95; 1.49\\
Spread/RMSE, $X$; $Y$; $Z$ & 2.77 & 0.47; 0.13; 0.14\\
$L_b$, $k$=1--5, 6--9, 10--18 & 22.6, 31.8, 38.2 & 7.8, 23.4, 17.4\\
$\bar q_b(0)$, same bands & 0.003, 0.003, 0.004 & 0.061, 0.029, 0.061\\
$\bar q(0)$, $X$; $Y$; $Z$ & 0.003 & 0.046; 0.56; 0.48\\
Mean ascent (MTU) & 0.29--1.20 & 0.15--2.14\\
\bottomrule
\end{tabular}
\end{table}

The assimilation constrains the slow level far better than the fast
levels. In Run C the analysis errors of $Y$ and $Z$ are comparable to
their climatological standard deviations, so forecasts begin with the
fast levels about half saturated ($\bar q_Y=0.56$, $\bar q_Z=0.48$)
and the slow level at $\bar q_X=0.05$; the ensemble is markedly
underdispersive. In Run U the $X$ analyses are far more accurate, with
initial normalized error 0.003. This measured asymmetry defines the
predictability problem studied here: how error in the slow level grows when the fast levels are
already largely uncertain.

\subsection{Scale-resolved error observables}\label{sec:observables}
Both regimes are analyzed through three Fourier bands of $X$:
$k=1$--5, 6--9 and 10--18, each including conjugate partners and
excluding the zero mode (nonzero modes below the Nyquist wavenumber 18
have weight two, and the Nyquist mode weight one). With band projection
$\Pi_b$, the error of launch $j$ and the fixed reference are
\begin{equation}
 E_b^{(j)}=\frac{\|\Pi_b(X_f^{(j)}-X_t^{(j)})\|^2}{K},\qquad
 L_b=\frac2K\left\langle\|\Pi_bX_t\|^2\right\rangle,
 \label{eq:energy}
\end{equation}
where the angle brackets denote an average over the 200-MTU truth run.
The references are
computed independently of the forecasts, not fitted. Nonzero Fourier
bands have zero spatial mean; their finite-record temporal mean need
not vanish. Under statistical homogeneity and negligible temporal band
means, $L_b$ approximates the mean squared difference between two
independent truth states. It is a saturation reference rather than an
exact upper bound; individual errors may exceed $L_b$.

Mean-curve diagnostics first average $q_b^{(j)}=E_b^{(j)}/L_b$ over all
200 launches at each lead and then take $\eta_b=\logit(\bar q_b)$. The
mean-ascent interval of each regime begins at the first lead at which
every band mean reaches 0.05 and ends before any band mean first
reaches 0.95, without re-entry; this gives 0.29--1.20~MTU for Run U and
0.15--2.14~MTU for Run C. Ordinary least squares fits $a_b+\hat r_bt$
on these intervals.

\subsection{Forecast-error energy budget}\label{sec:errorbudget}
Because truth and forecasts obey Eq.~\eqref{eq:osse}, the error
$\delta X=X_f-X_t$ (and likewise $\delta Y$, $\delta Z$) satisfies an
exact equation. Let $\B(a,b)_k=a_{k-1}(b_{k+1}-b_{k-2})$ denote the
advection on the $X$ ring, so that the $X$ advection term is $\B(X,X)$,
and let $(\Sop Y)_k=\sum_jY_{kJ+j}$. Then
\begin{equation}
 \delta\dot X=\B(X_t,\delta X)+\B(\delta X,X_t)+\B(\delta X,\delta X)
 -\delta X-\alpha\Sop\delta Y ,
 \label{eq:errdyn}
\end{equation}
and the band energies of Eq.~\eqref{eq:energy} obey
\begin{equation}
 \dot E_b=\mathcal P_b+\mathcal N_b+\mathcal C_b-2E_b,
 \label{eq:bandbudget}
\end{equation}
with truth-referenced production, error--error self-advection
redistribution and interlevel exchange,
\begin{equation}
 \begin{aligned}
 \mathcal P_b&=\tfrac2K\langle\Pi_b\delta X,\,
   \B(X_t,\delta X)+\B(\delta X,X_t)\rangle,\\
 \mathcal N_b&=\tfrac2K\langle\Pi_b\delta X,\,\B(\delta X,\delta X)\rangle,\\
 \mathcal C_b&=-\tfrac{2\alpha}K\langle\Pi_b\delta X,\,\Sop\delta Y\rangle.
 \end{aligned}
 \label{eq:bandterms}
\end{equation}
Because $\langle x,\B(x,x)\rangle=0$ on the periodic ring, the triad
terms sum to zero over the complete Fourier partition (including $k=0$):
$\mathcal N_b$ redistributes error energy among scales without creating
it. This is the conservative part of the truth-referenced split, not
the complete cross-band interaction: $\mathcal P_b$ also contains
truth--error interactions among bands. With midpoint reference
$M=(X_f+X_t)/2$, the advection difference is instead
$\B(M,\delta X)+\B(\delta X,M)$, absorbing the explicit
error--error term. Thus the split into $\mathcal P_b$ and
$\mathcal N_b$ depends on the reference, while their sum and the
interlevel exchange do not. In Run U, $\mathcal C_b\equiv0$.

At the level scale, the unnormalized error energies
$\mathcal E_X=\tfrac12\|\delta X\|^2$, $\mathcal E_Y=\tfrac12\|\delta Y\|^2$
and $\mathcal E_Z=\tfrac12\|\delta Z\|^2$ obey
\begin{equation}
 \begin{aligned}
 \dot{\mathcal E}_X&=\mathcal P_X-2\mathcal E_X-\mathcal T^{\varepsilon}_{XY},\\
 \dot{\mathcal E}_Y&=\mathcal P_Y-2c_x\mathcal E_Y+\mathcal T^{\varepsilon}_{XY}
   -\mathcal T^{\varepsilon}_{YZ},\\
 \dot{\mathcal E}_Z&=\mathcal P_Z-2c_y\mathcal E_Z+\mathcal T^{\varepsilon}_{YZ},
 \end{aligned}
 \label{eq:levelbudget}
\end{equation}
where $\mathcal P_{X,Y,Z}$ are the productions by the linearized
advection of each ring (the error--error advection vanishes on each
ring) and
\begin{equation}
 \mathcal T^{\varepsilon}_{XY}=\alpha\sum_m\delta X_{k(m)}\,\delta Y_m,\qquad
 \mathcal T^{\varepsilon}_{YZ}=\beta\sum_n\delta Y_{m(n)}\,\delta Z_n .
 \label{eq:errortransfer}
\end{equation}
The interlevel error transfers cancel pairwise in the total error energy
and are therefore unambiguous exchanges: positive values denote
$X\to Y$ and $Y\to Z$ transfer, respectively, and negative values
the reverse. They can take either sign, as can the state-energy
transfers in principle, although those are positive at every sampled
truth state of Run C. Summed over the complete Fourier partition,
including $k=0$, $\sum_b\mathcal C_b=-(2/K)\mathcal
T^{\varepsilon}_{XY}$. The OSSE output stores the analysis mean at every
assimilation cycle and the truth at every output step, so the complete
error fields, and every term in
Eqs.~\eqref{eq:bandbudget}--\eqref{eq:errortransfer}, can be regenerated
by re-integrating the forecasts from their analyses.

\subsection{Correlation diagnostics}\label{sec:diagnostics}
Three correlation diagnostics address different questions. The first
correlates the mean logits $\eta_b$ over lead time within the
mean-ascent interval (shared timing). The second correlates their
residuals after subtracting each band's affine fit (coordinated
departures from constant-rate timing). The third uses individual logits
$\ell_{bj}(t)=\logit(q_b^{(j)}(t))$: at each lead, the three bands are
centered and standardized across launches, the three-band Pearson matrix
is formed, and these matrices are averaged with equal weight over leads.
This removes the shared mean lead dependence and measures
launch-to-launch alignment of band errors. The primary interval for the
third diagnostic, 0--0.26~MTU (27 leads), retains all 200 launches in both
regimes because no individual band reaches 0.95 before 0.26~MTU; no
clipping or survivor selection is involved. Later-interval estimates use,
at each lead, the launches with $0.05<q_b^{(j)}<0.95$ in all three
bands, and are reported with their coverage.

As an alignment reference, whole launch trajectories are permuted
independently among bands 1000 times. This preserves each band's
marginal mean curve, and hence the first two diagnostics, while
destroying joint launch alignment. Because launches overlap in time,
the permutation distribution is a descriptive reference rather than an
exchangeability-based significance test.

Sampling uncertainty is estimated by 2000 circular moving-block
bootstrap replicates of whole launch trajectories, keeping bands and
leads together. The primary block length is five launches, with three
and ten as sensitivities; six-MTU threshold diagnostics also use fifteen.
The C initial-error experiments use paired launch resamples. We report
90\% pointwise percentile intervals conditional on fixed climatological
references and declared lead windows, refitting rates and frames in each
replicate. Undefined logits or missing threshold crossings are counted
rather than clipped; intervals for such metrics condition additionally
on the defined replicates. A separate fixed 0.1--0.9 mean-ascent window checks
endpoint sensitivity. These intervals assess variability within each
archive; they do not establish independent climatological replication.

%% ======================================================================
\section{Results}\label{sec:results}
\subsection{Growth clocks without interlevel coupling}
\begin{figure*}[t]
\includegraphics[width=0.96\textwidth]{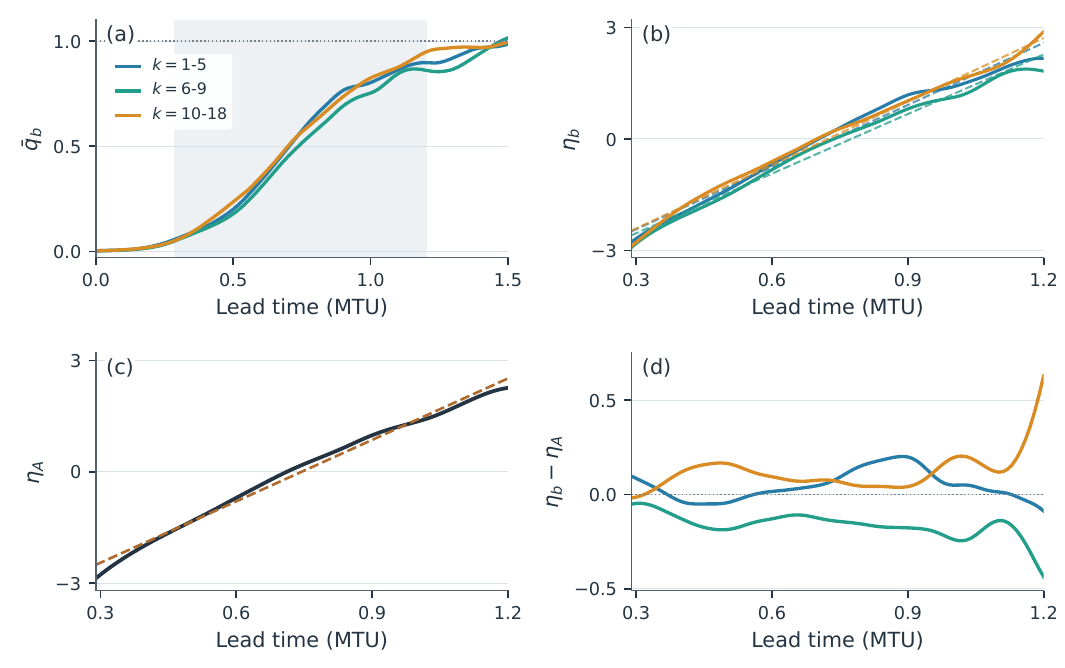}
\caption{Mean error growth and scale-resolved logit clocks in Run U
(interlevel-uncoupled), from 200 forecasts. The three Fourier bands of
the 36-variable $X$ ring, $k=1$--5, 6--9 and 10--18, interact through
the quadratic advection. (a) Band-mean normalized errors
$\bar q_b=\langle E_b\rangle/L_b$ with fixed references $L_b$
[Eq.~\eqref{eq:energy}]. Shading marks the mean-ascent interval, from
the first lead at which all band means reach 0.05 to the last lead
before any reaches 0.95 (0.29--1.20~MTU, 92 leads). (b) Band logits
$\eta_b=\logit(\bar q_b)$ with least-squares affine fits (dashed);
$R^2=0.985$, 0.989 and 0.993 in order of increasing wavenumber.
(c) Logit of the reference-weighted aggregate
$A=\sum_bL_b\bar q_b/\sum_bL_b$ with its affine fit ($R^2=0.991$).
(d) Band-minus-aggregate contrasts $\eta_b-\eta_A$; zero denotes equal
normalized band and aggregate errors. All 200 forecasts are averaged at
every lead before the logit is taken; no smoothing, clipping or case
selection is applied.}
\label{fig:runU}
\end{figure*}

Figure~\ref{fig:runU} shows the mean error growth in Run U. The three
band means rise together from $\bar q_b\approx0.003$ at launch and
approach their references near 1.5~MTU. Over the ascent interval the
band logits are nearly affine, with $R^2=0.985$, 0.989 and 0.993 and
rates $\hat r_b=5.57$, 5.35 and 5.70~MTU$^{-1}$ in order of increasing
wavenumber, and the aggregate logit has $R^2=0.991$
(rate 5.51~MTU$^{-1}$). The band-minus-aggregate contrasts remain within
$\pm0.2$ of zero for most of the interval, with the middle band
persistently below the aggregate, and diverge only near the end of the
ascent as the high band approaches saturation. The bands are Fourier
subsets of one nonlinear ring, so Run U is not a realization of three
independent logistic bands; the exact benchmark of
Sec.~\ref{sec:geometry} supplies that reference.

\subsection{Growth clocks with coupling}
\begin{figure*}[t]
\includegraphics[width=0.96\textwidth]{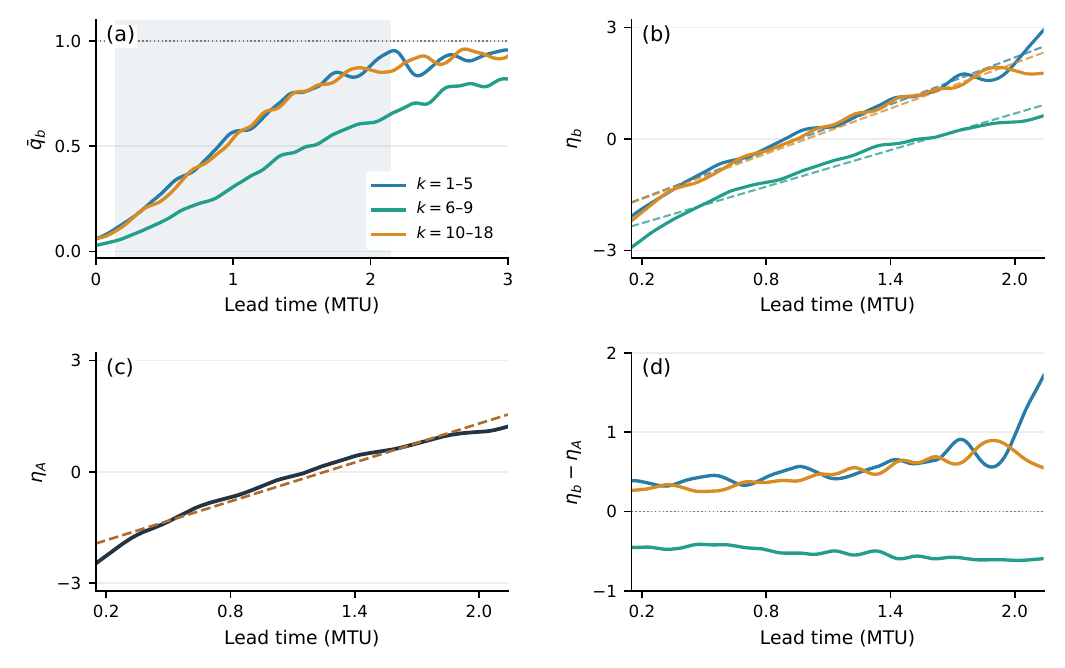}
\caption{As in Fig.~\ref{fig:runU}, for the coupled Run C.
(a) Normalized band-mean errors; shading marks the mean-ascent interval,
0.15--2.14~MTU (200 leads), defined by the same 0.05/0.95 rule.
(b) Band logits and affine fits, with $R^2=0.983$, 0.965 and 0.977 in
order of increasing wavenumber. (c) Aggregate logit and affine fit
($R^2=0.971$). (d) Band-minus-aggregate contrasts; the middle band
remains below the aggregate throughout the ascent, while
the mean clocks remain nearly affine. Horizontal and contrast-axis
ranges differ from Fig.~\ref{fig:runU} to accommodate the longer
coupled ascent.}
\label{fig:runC}
\end{figure*}

With coupling, the same observables give slower and more differentiated
mean clocks (Fig.~\ref{fig:runC}). The band rates are 2.11, 1.64 and
2.03~MTU$^{-1}$, lower than in Run U by factors of 2.6, 3.3 and 2.8, and
the aggregate rate is 1.74~MTU$^{-1}$. The logits remain nearly affine,
with $R^2=0.983$, 0.965 and 0.977 and aggregate $R^2=0.971$. The middle
band lies below the aggregate with a logit contrast of about
$-0.5$ throughout the ascent, while the low and high bands advance
together. The coupled regime also differs in its initial condition:
forecasts begin with band errors an order of magnitude larger than in
Run U and with the fast levels about half saturated
(Table~\ref{tab:config}). Over a common elapsed interval of
0.29--1.20~MTU the coupled rates are 2.23, 1.96 and 2.33~MTU$^{-1}$
($R^2=0.983$, 0.983 and 0.995), so the reduction of the clock rates
relative to Run U does not depend on the choice of interval. The 90\%
block-bootstrap intervals of the band rates are also well separated:
5.17--6.22, 4.94--5.87 and 5.23--6.34~MTU$^{-1}$ in Run U, and
1.78--2.48, 1.51--1.78 and 1.82--2.34~MTU$^{-1}$ in Run C (the Run C
low-band interval is conditional on the 88\% of replicates in which its
mean logit is defined; Sec.~\ref{sec:geometrysec}).

\subsection{How distinctive is the logit clock?}\label{sec:distinctive}
The logit supplies a physically normalized reference clock even when
several coordinates straighten a mean ascent. Table~\ref{tab:transforms}
compares affine fits of four clock coordinates over the same intervals; by
Eq.~\eqref{eq:etaLambda} and the discussion following it, each
coordinate corresponds to a distinct growth-rate profile $\Lambda(q)$.
The exponential coordinate $\ln\bar q_b$ is less affine over these
ascent intervals ($R^2=0.870$--0.909). The other three coordinates all give
$R^2>0.965$. The Gompertz coordinate yields the largest $R^2$ for five
of the eight curves, the logit for two, and the untransformed error for
one.

\begin{table*}[t]
\caption{Coefficient of determination $R^2$ of least-squares affine fits
in lead time of alternative clock coordinates of the band-mean
normalized errors $\bar q_b$ and of the aggregate $A$, over the
mean-ascent intervals of Figs.~\ref{fig:runU} and~\ref{fig:runC}. The
second column gives the growth-rate profile $\Lambda=\dd\ln q/\dd t$ for
which each coordinate advances at constant speed
[Eq.~\eqref{eq:etaLambda}]. Values measure linearity in each
coordinate's own metric, not predictive skill under a common error-energy
loss. Bold values mark the largest transformed $R^2$ in each column.}
\label{tab:transforms}
\small
\begin{tabular}{llcccccccc}
\toprule
 & & \multicolumn{4}{c}{Run U, 0.29--1.20 MTU} & \multicolumn{4}{c}{Run C, 0.15--2.14 MTU}\\
\cmidrule(lr){3-6}\cmidrule(lr){7-10}
Coordinate & Constant-speed $\Lambda(q)$ & $k$=1--5 & 6--9 & 10--18 & $A$ & 1--5 & 6--9 & 10--18 & $A$\\
\midrule
$q$ (constant energy rate) & $c/q$ & 0.971 & 0.985 & 0.988 & 0.984 & 0.979 & \textbf{0.995} & 0.969 & 0.986\\
$\ln q$ (exponential) & $\lambda$ & 0.884 & 0.900 & 0.879 & 0.888 & 0.875 & 0.909 & 0.870 & 0.889\\
$\logit q$ (logistic) & $\lambda(1-q)$ & 0.985 & 0.989 & \textbf{0.993} & 0.991 & \textbf{0.983} & 0.965 & 0.977 & 0.971\\
$-\ln(-\ln q)$ (Gompertz) & $-\lambda\ln q$ & \textbf{0.994} & \textbf{0.996} & 0.991 & \textbf{0.997} & 0.981 & 0.994 & \textbf{0.990} & \textbf{0.995}\\
\bottomrule
\end{tabular}
\end{table*}

These values describe coordinate linearity, not the physical cause of
an ascent or a common-loss ranking of growth laws. In particular,
Gompertz-like behavior is a plausible consequence of unresolved
heterogeneity in growth onset, rate or saturation stage, but this table
does not test that mechanism. Run U rises through a sigmoidal mean
ascent, while the slow-level mean errors of Run C are nearly linear in
time over much of the selected interval. A linear error-energy profile
is compatible with propagation across active scales, but its shape
alone does not identify a signed upscale transfer.

The logit's contribution is its exact logistic reference geometry:
fixed physical normalization, additive clock departures, an explicit
inverse, and a common frame for the physical budget. As noted in
Sec.~\ref{sec:theory}, a Gompertz reference would support the same
geometry with a different denominator. The signed exchanges and
threshold delays reported below do not depend on the clock coordinate;
the magnitudes of transformed contributions do, and are interpreted
relative to the declared logistic reference. Sections~\ref{sec:exchange}
and~\ref{sec:attribution} use measured tendencies and
initial-condition partitions to interpret departures dynamically.

\subsection{Shared timing and launch-to-launch dependence}
\begin{figure*}[t]
\includegraphics[width=0.96\textwidth]{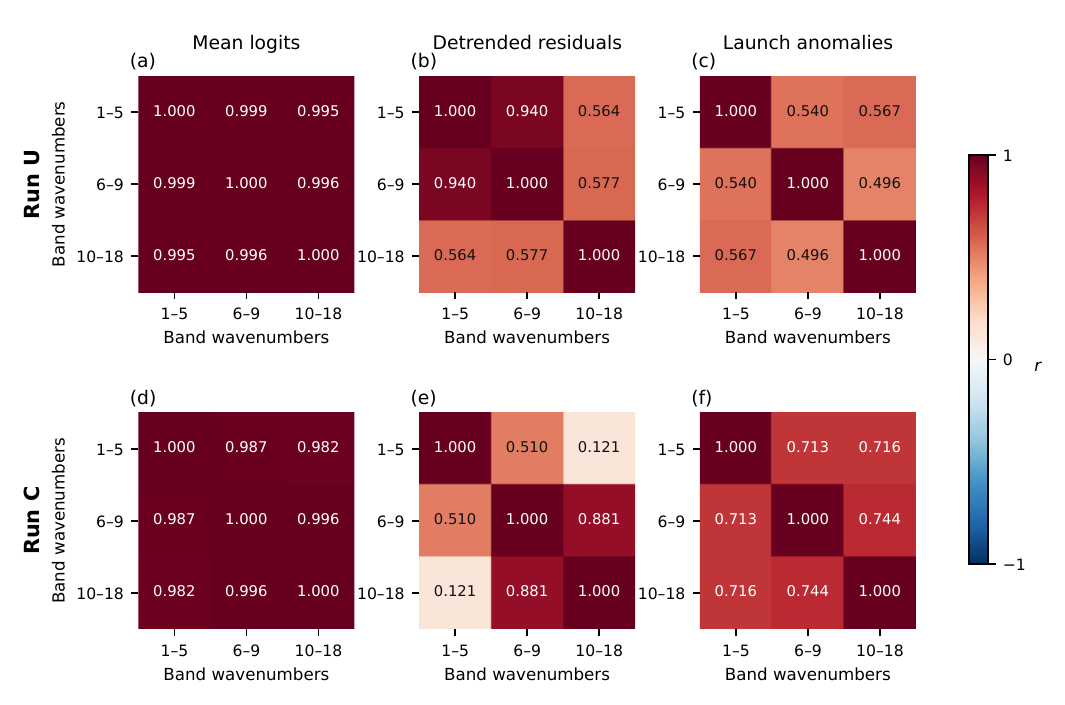}
\caption{Three correlation diagnostics of the band logits. Upper row
(a--c): Run U; lower row (d--f): Run C. (a,d) Temporal correlations of
the logits of the 200-launch mean errors over each run's mean-ascent
interval. (b,e) Correlations of the residuals after subtracting each
band's affine clock. (c,f) Launch-to-launch correlations of individual
logits: at each lead the logits are centered and standardized across
launches, and the per-lead correlation matrices are averaged over the
27 leads 0--0.26~MTU, using all 200 launches without clipping or
selection. The columns thus measure shared timing, coordinated
departures from constant-rate timing, and alignment of forecast-to-forecast
variations, respectively.}
\label{fig:correlations}
\end{figure*}

Figure~\ref{fig:correlations} separates the three correlation
questions. In Run U the mean logits are almost perfectly correlated in
time (0.995--0.999), as expected from a dominant shared growth direction
[Eq.~\eqref{eq:covariance}]. After each affine clock is subtracted, the
residuals remain strongly correlated for the low and middle bands (0.940)
but only moderately for pairs involving the high band (0.564 and 0.577).
These residuals describe coordinated departures from constant-rate
timing, which can reflect common nonlinear timing as well as motion
across the reference direction. After centering and standardizing
individual logits at each lead, the launch-to-launch anomalies of
different bands remain positively correlated (0.496--0.567); the
leading eigenvalue of the averaged correlation matrix accounts for 69.0\%
of the standardized variance. Shared mean lead dependence cannot produce
this alignment: permuting whole launch trajectories among bands leaves
the first two diagnostics unchanged and moves every off-diagonal median
of the third to within 0.005 of zero in both runs.

The coupled regime has a different structure. Its mean logits are again
highly correlated (0.982--0.996), but the detrended residuals separate
the low band from the others (0.510 and 0.121) while linking the middle
and high bands (0.881). The early launch-to-launch correlations are
stronger than in Run U (0.713--0.744), with a leading eigenvalue fraction
of 81.6\%. This difference is present at initialization: Run C forecasts
begin more aligned across bands, whereas the alignment in Run U grows
more rapidly during the early interval, and the ordering is not uniform
for every pair at every lead. Later in the ascent, the interior
populations shrink (34--162 launches per lead in Run U and 55--150 in
Run C) and the conditional correlations are weaker and similar in the
two regimes (0.290--0.341 and 0.300--0.392). Growth stage and population
are therefore part of the interpretation, and the data do not support a
single ordering of alignment by coupling strength. The strong mean-clock
correlations in both regimes illustrate the central point of
Eq.~\eqref{eq:covariance}: a highly correlated mean ascent can coexist
with substantial and differently organized launch-to-launch variation.

The leading eigenvalue fractions have 90\% block-bootstrap intervals
of 0.653--0.724 in Run U and 0.781--0.842 in Run C, compared with
point estimates 0.690 and 0.816. These early-interval estimates are
defined in every replicate and remain separated in the tested
block-length sensitivities. They describe distinct initial-error
alignment in the two archives, not an isolated causal effect of coupling.

\subsection{Longitudinal and transverse evolution}\label{sec:geometrysec}
\begin{figure*}[t]
\includegraphics[width=0.96\textwidth]{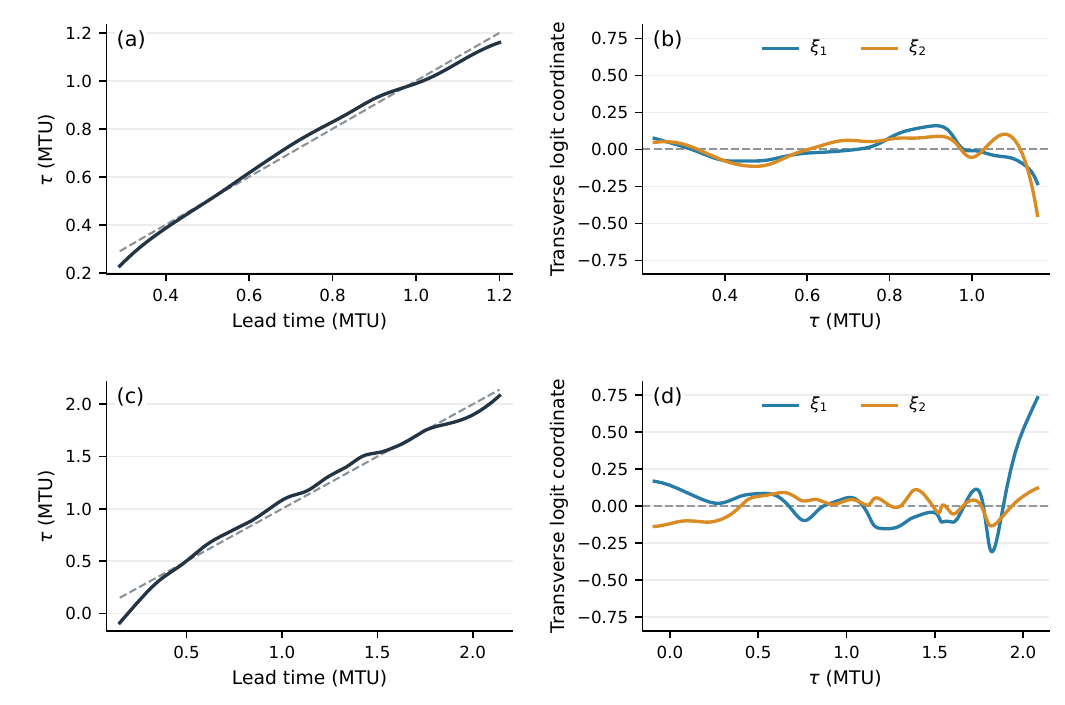}
\caption{Longitudinal clocks and transverse geometry of the mean
trajectories for Run U (a,b) and Run C (c,d), over their mean-ascent
intervals and relative to the affine fits of Figs.~\ref{fig:runU}
and~\ref{fig:runC}. (a,c) Longitudinal coordinate $\tau$ versus lead
time; the dashed line $\tau=t$ is the exact affine-clock benchmark.
(b,d) The two orthonormal transverse coordinates of
Eq.~\eqref{eq:transversebasis} versus $\tau$; zero marks the reference
line. The decomposition reconstructs the band logits exactly as
$\boldsymbol\eta=\boldsymbol a+\boldsymbol v\tau+\boldsymbol w_1\xi_1+
\boldsymbol w_2\xi_2$. An extrapolated fitted origin can yield $\tau<0$
at early leads; $\tau$ is a coordinate, not a forecast time.}
\label{fig:clocktransverse}
\end{figure*}

Figure~\ref{fig:clocktransverse} applies Eq.~\eqref{eq:geometry} to each
mean trajectory, with $\boldsymbol v=\hat{\boldsymbol r}$ and
$\boldsymbol a$ taken from the affine fits of Figs.~\ref{fig:runU}
and~\ref{fig:runC}. For three bands we use the orthonormal transverse
basis
\begin{equation}
 \boldsymbol w_1=\frac{\boldsymbol e_1-u_1\boldsymbol u}
                         {\sqrt{1-u_1^2}},\qquad
 \boldsymbol w_2=\boldsymbol u\times\boldsymbol w_1,
 \label{eq:transversebasis}
\end{equation}
with $\boldsymbol e_1=(1,0,0)^{\mathsf T}$ and
$\xi_i=\boldsymbol w_i^{\mathsf T}(\boldsymbol\eta-\boldsymbol a)$;
both empirical rate vectors satisfy $|u_1|<1$, and the orientation is a
coordinate convention. With the fitted rates,
$\boldsymbol w_1\approx(0.81,-0.40,-0.42)^{\mathsf T}$ and
$\boldsymbol w_2\approx(0,0.73,-0.68)^{\mathsf T}$ in Run U, and
$\boldsymbol w_1\approx(0.78,-0.40,-0.49)^{\mathsf T}$ and
$\boldsymbol w_2\approx(0,0.78,-0.63)^{\mathsf T}$ in Run C. Thus
$\xi_1$ measures primarily the low band relative to the other two, and
$\xi_2$ the middle band relative to the high band. The longitudinal coordinate remains close to
elapsed lead time: the root-mean-square difference $\tau-t$ is 0.025~MTU
in Run U and 0.072~MTU in Run C, or 2.7\% and 3.6\% of the respective
ascent durations. The root-mean-square transverse norms are 0.124 and
0.179 logit units. The longitudinal speed exceeds one at the beginning of both
intervals. Near the end it falls below one in Run U but rises above one
again in Run C; departures from the fitted clock need not be monotone.
Transverse excursions are more pronounced near the edges of the chosen
intervals. All plotted band means remain below 0.95, so these excursions
are interior departures rather than mean-error overshoots. Subtracting
separate affine clocks, forming band-minus-aggregate contrasts and
projecting transversely are therefore related but distinct diagnostics:
only the last isolates departures from the reference band-growth
relation from changes in the common clock speed.

The endpoint sensitivity of a mean-logit clock is visible in the
bootstrap: 11.85\% of primary resamples put Run C's low-band mean outside
$(0,1)$ somewhere in the original ascent interval. Restricting the fixed
window to the 0.1--0.9 mean ascent reduces this fraction to 0.8\%.
We retain the original-window point estimates, while rate and frame
intervals with undefined replicates are conditional on the defined
population. This motivates using declared interior windows and retaining
untransformed budget diagnostics alongside the clocks.

\subsection{Direction of forecast-error energy exchange between slow and fast levels}\label{sec:exchange}
We re-integrated all 200 forecasts in each regime from their stored
analysis means. Maximum relative differences from archived band errors
are $1.2\times10^{-15}$ in Run U over all six MTU and
$1.6\times10^{-10}$ in Run C through its 2.14-MTU ascent. Direct
instantaneous budget identities close to $7\times10^{-14}$ in scaled
error, and conservative error--error terms cancel across the complete
Fourier partition to $3\times10^{-15}$. Bandwise centered-difference
discrepancies at 0.01-MTU output spacing are below 1.7\% over the control
ascents, using mean absolute discrepancy divided by mean absolute
finite-difference tendency. The faster $Y$ level of Run C requires a
finer sampling check: its 5.9\% discrepancy at 0.01-MTU spacing falls to
0.06\% at 0.001-MTU spacing in a short matched-lead check near 1~MTU.
This distinguishes numerical sampling from the algebraic budget.
Individual coupled trajectories diverge from their archived versions
at late leads through chaotic amplification of round-off; at 6~MTU the
maximum relative band-error difference is 15\%, while each band-mean
difference remains below 0.09\%. The replay is therefore numerically
faithful over the ascent, with late ensemble comparisons interpreted
at their stated precision.

\begin{figure*}[t]
\centering
\includegraphics[width=0.86\textwidth]{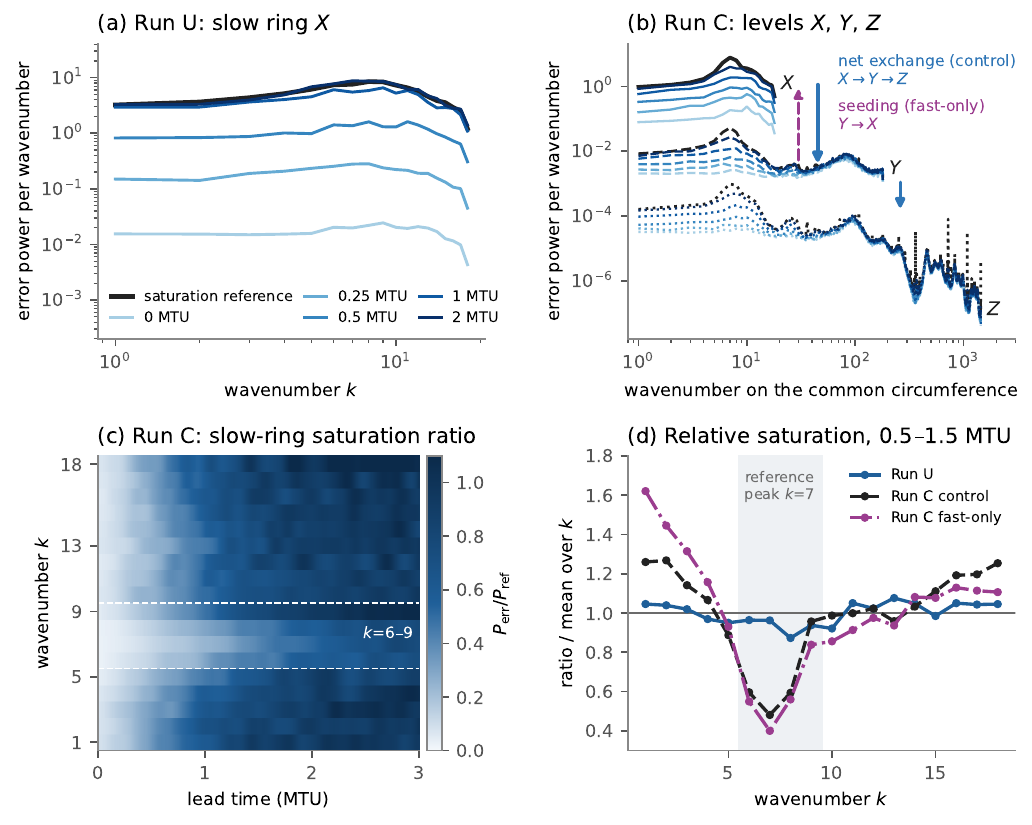}
\caption{Error spectra and saturation ratios from replays of the 200
forecasts. (a)~Run U: launch-mean slow-ring error power per wavenumber at
leads 0, 0.25, 0.5, 1 and 2~MTU (light to dark) and the saturation
reference $2\langle P_k\rangle$ (black). (b)~Run C: the same for the $X$
(solid), $Y$ (dashed) and $Z$ (dotted) rings, with the $Y$ and $Z$ rings
mapped onto the circumference of the $X$ ring. Arrows indicate the
measured direction of net interlevel exchange in the control and of
seeding in the fast-only experiment (Sec.~\ref{sec:exchange}).
(c)~Run C: slow-ring saturation ratio $P_{\rm err}/P_{\rm ref}$ against
lead and wavenumber; dashed lines bound $k=6$--9. (d)~Saturation ratio
divided by its wavenumber mean, averaged over 0.5--1.5~MTU, for Run U,
the Run C control and the Run C fast-only experiment. The replays
reproduce the archived band errors to $10^{-15}$ (Run U) and
$1.5\times10^{-10}$ (Run C, through 2.14~MTU).}
\label{fig:spectra}
\end{figure*}

Figure~\ref{fig:spectra} shows the launch-mean error power of the
replayed forecasts per wavenumber, normalized so that its sums over a
band give $E_b$ of Eq.~\eqref{eq:energy}, against saturation spectra
$2\langle P_k\rangle$ computed from the truth run, the per-wavenumber
analogue of $L_b$. In Run U the slow-ring error spectrum grows nearly
parallel to its saturation spectrum: there is no front moving from high
to low wavenumbers, and the band clocks advance together. In Run C the
fast levels begin about half saturated, at energy-weighted ratios of
0.56 for $Y$ and 0.48 for $Z$, whereas the slow ring begins at 0.05
(Table~\ref{tab:config}). This is the ordering expected when small, fast
scales saturate first. The budget below shows, however, that the net
error-energy exchange between levels is directed from slow to fast.

\begin{figure*}[t]
\centering
\includegraphics[width=0.98\textwidth]{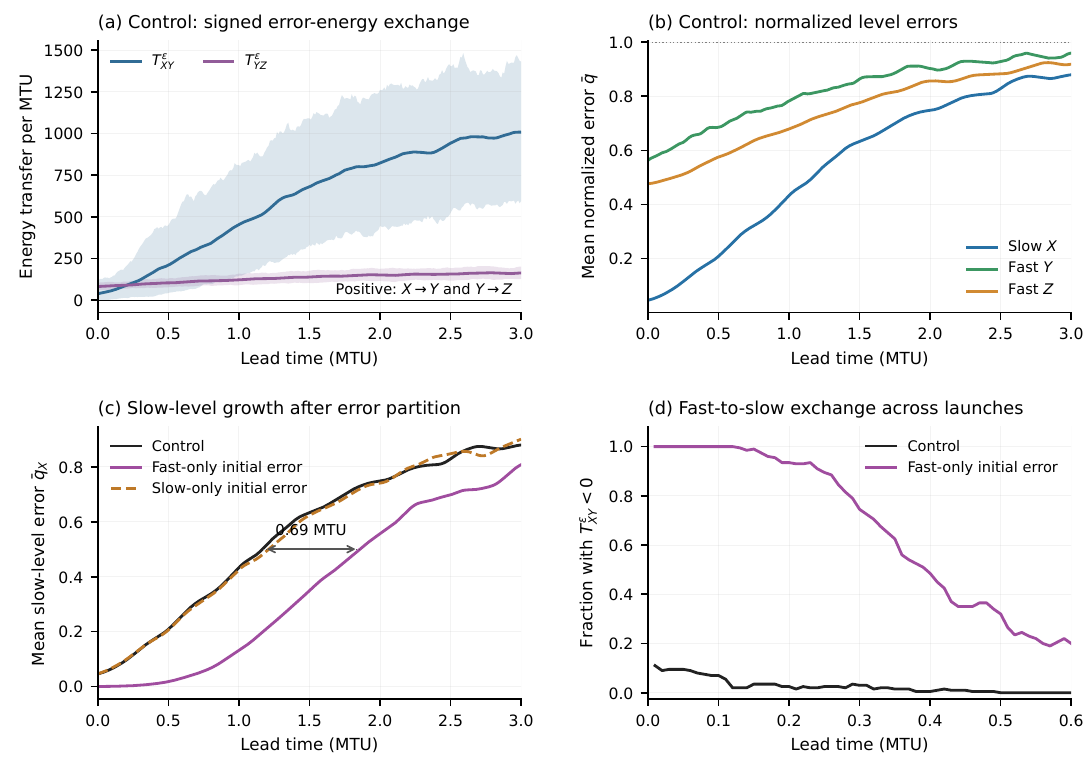}
\caption{Error-energy exchange and initial-error partitions in Run C,
using all 200 launches. (a) Control interlevel transfers; lines show
launch means and shading the 10th--90th percentiles across launches,
not confidence intervals. Positive values denote $X\to Y$ and
$Y\to Z$ exchange, respectively. (b) Control mean errors normalized by the fixed level
references. (c) Slow-level mean error for the control, fast-only initial
error (true $X$, analyzed $Y,Z$), and slow-only initial error (analyzed
$X$, true $Y,Z$). The arrow marks the equal-error delay at
$\bar q_X=0.5$. (d) Fraction of launches with fast-to-slow exchange in
the control and fast-only experiments; lead zero is excluded because
the latter has exactly zero initial exchange. Panels (a--c) cover
0--3~MTU; panel (d) resolves the early reversal.}
\label{fig:exchange}
\end{figure*}

The control exchanges are predominantly directed from $X$ to $Y$
and from $Y$ to $Z$ (Fig.~\ref{fig:exchange}a,b). The fractions of launch--lead pairs
with $\mathcal T^\varepsilon_{XY}>0$ are 91.3\%, 97.3\% and 99.8\% in
$(0,0.1]$, $(0.1,0.3]$ and $(0.3,2.2]$~MTU, respectively; the mean
$\mathcal T^\varepsilon_{YZ}$ is also positive in each interval.
Integrated over the C ascent, $\mathcal C_b$ removes 34.6\%, 35.9\% and
24.9\% of the corresponding integrated $\mathcal P_b$ in increasing
band order. These ratios describe signed terms in the recorded budget,
not an intervention that turns coupling off. Their 90\% block-bootstrap
intervals are 31.6--38.1\%, 33.1--39.4\% and 23.1--26.7\%,
respectively, when expressed as positive removal magnitudes.
At the level scale, the launch-mean ratio
$\overline{\mathcal T^\varepsilon_{XY}}/\overline{\mathcal E}_X$ stays
between 1.01 and 1.28~MTU$^{-1}$ over the ascent. In the slow-level
budget of Eq.~\eqref{eq:levelbudget} the net exchange therefore acts
like an additional, nearly steady damping of slow-level error, about
60\% as large as the explicit damping $2\mathcal E_X$ and 31\% of the
slow-level production in time integral. This is a descriptive rate
along the recorded trajectories, not a closure. A plausible
interpretation is that the fast levels respond coherently to the
slow-level error that forces them, while their incoherent initial errors
partly cancel in the sum $\Sop\delta Y$.

Fast-level errors nevertheless seed slow-level errors. Replacing only
the initial $X$ field by truth gives the fast-only experiment; replacing
initial $Y,Z$ by truth gives the slow-only experiment. For fast-only
initial error, Eq.~\eqref{eq:errdyn} implies
\begin{equation}
 \begin{aligned}
 \delta X(t)&=-\alpha\Sop\delta Y(0)t+O(t^2),\\
 \mathcal T^\varepsilon_{XY}(t)
   &=-\alpha^2\|\Sop\delta Y(0)\|^2t+O(t^2).
 \end{aligned}
 \label{eq:fastseeding}
\end{equation}
Thus the influence begins immediately, with initially quadratic
slow-error energy growth. All sampled launch--lead pairs have reversed
exchange in $(0,0.1]$~MTU, and 92.1\% do in $(0.1,0.3]$~MTU; the
fraction falls to 8.3\% in $(0.3,2.2]$~MTU as the net exchange becomes
slow-to-fast. The fast-only mean $\bar q_X$ reaches 0.132 at 1~MTU and
0.556 at 2~MTU, compared with control values 0.433 and 0.748.

Figure~\ref{fig:exchange}c quantifies the subsequent timing difference.
The fast-only mean reaches $\bar q_X=0.25,0.5,0.7$ later than the control
by 0.704, 0.690 and 0.762~MTU, respectively
(Table~\ref{tab:thresholddelay}). These are equal-error threshold delays,
not delays before influence begins. The delay is nearly independent of
the threshold: for every $0.13\le\bar q_X\le0.70$ it lies between 0.62
and 0.76~MTU, within the bootstrap intervals of
Table~\ref{tab:thresholddelay}. After the initial seeding transient, the
fast-only mean therefore approximately follows a time-shifted control
growth curve over the tested error range: a shift of 0.70~MTU aligns the
two mean curves with a root-mean-square difference of 0.009 in $\bar q_X$
for $0.13\le\bar q_X\le0.70$, and at 1~MTU the fast-only mean has the
error the control had at 0.29~MTU. This is a scalar offset of the
slow-level growth clock. It does not establish equal growth speeds at
matched error, which differ locally in finite differences of the saved
means, nor an unchanged band composition. At $\bar q_X=0.5$, for example,
the low-band logit is 0.34 higher and the high-band logit 0.15 lower in
the fast-only experiment, consistent with its different relative
spectrum (Fig.~\ref{fig:spectra}d). Similar total-error histories thus
accompany different spectral evolution, as anticipated in
Sec.~\ref{sec:severalclocks}.
The fast-only curve first reaches
90\% of the simultaneous control mean at 2.94~MTU. In contrast, retaining
only the archived slow-level initial error keeps the mean $X$-error
curve within 4.71\% of the control over 0.15--2.14~MTU. These partition
experiments concern the actual archived initial-error distributions;
the slow and fast perturbations are not amplitude matched, and the
responses do not define additive causal fractions.

\begin{table}[t]
\caption{Delay in first attainment of a slow-level mean error threshold
for fast-only initial error relative to the control. Crossings are
linearly interpolated between saved leads. The 90\% percentile intervals
use 2000 paired circular block resamples with block length five
launches; every threshold is attained in every replicate, including
block-length sensitivities three, ten and fifteen.}
\label{tab:thresholddelay}
\begin{tabular}{ccc}
\toprule
Threshold $\bar q_X$ & Delay (MTU) & 90\% interval (MTU)\\
\midrule
0.25 & 0.704 & [0.625, 0.777]\\
0.50 & 0.690 & [0.603, 0.786]\\
0.70 & 0.762 & [0.543, 1.024]\\
\bottomrule
\end{tabular}
\end{table}

\subsection{Budget attribution of clock speed and transverse motion}\label{sec:attribution}
\begin{figure*}[t]
\centering
\includegraphics[width=0.98\textwidth]{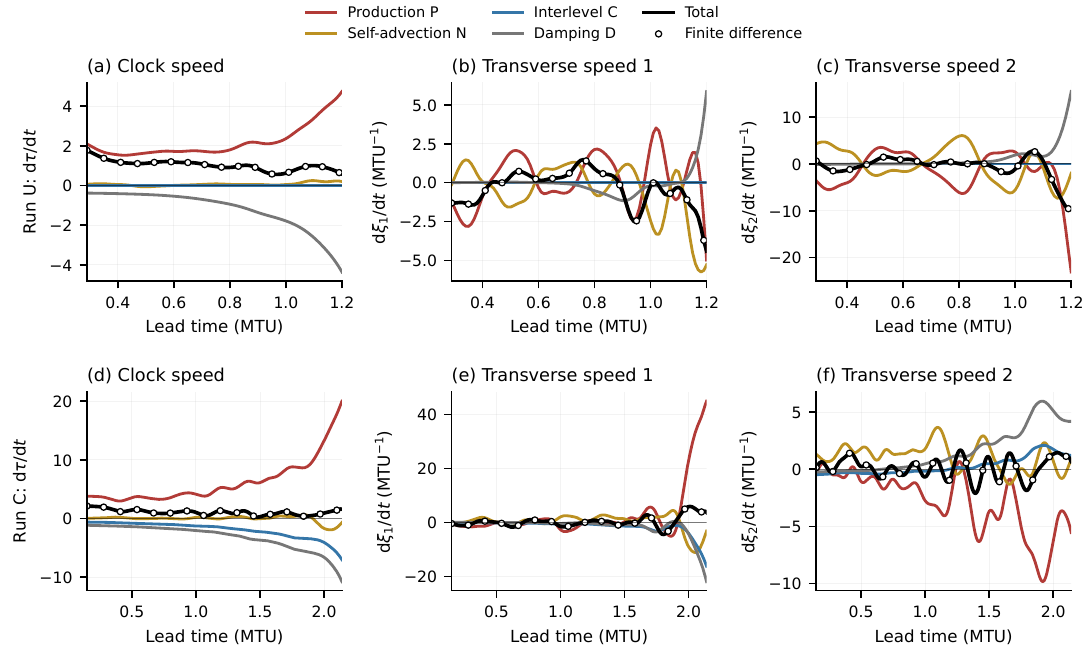}
\caption{Physical attribution of clock and transverse tendencies.
Rows show Run U (a--c) and Run C (d--f) on their original mean-ascent
intervals; columns show $\dot\tau$, $\dot\xi_1$ and $\dot\xi_2$.
Colored curves are the transformed contributions of truth-referenced
production $\mathcal P$, error--error redistribution $\mathcal N$,
interlevel exchange $\mathcal C$ and damping $D=-2E$. Black curves
show their sum; open circles show centered finite differences of the
coordinates at selected leads, using 0.01-MTU spacing and no smoothing.
All 200 launches are averaged before transformation. Each row uses its
own fitted frame. Opposing contributions need not lie between zero and
the net tendency.}
\label{fig:attribution}
\end{figure*}

Figure~\ref{fig:attribution} transforms the launch-mean physical budget
into the same frame as Fig.~\ref{fig:clocktransverse}. For
$S^{(m)}\in\{\mathcal P,\mathcal N,\mathcal C,-2E\}$, each contribution is
$\overline S{}_b^{(m)}/[L_b\bar q_b(1-\bar q_b)]$, followed by the
projections of Eq.~\eqref{eq:geometry}. Averaging precedes transformation.
The sum agrees with centered finite differences of the mean logits to
within 0.55\% in relative root-mean-square discrepancy, and with those
of $\tau,\xi_1,\xi_2$ to within 1.71\%, using the finite-difference RMS
as denominator and 0.01-MTU spacing.

Production advances the mean clock, while damping and, in Run C,
interlevel exchange oppose it (Table~\ref{tab:clockbudget}). Large
opposing contributions leave a net longitudinal speed near one in each
regime's fitted frame. These are signed contributions along recorded
trajectories. Since the frames are fitted separately, their values do
not assign a causal fraction of the U--C rate difference. Nor would
subtracting $\mathcal C$ from the recorded tendency generate an
uncoupled trajectory: the other terms depend on the coupled states.
Because $\tau$ is measured in units of each regime's fitted time, the
two columns are not directly comparable. In common logit units, the
band-averaged transformed production is similar in the two regimes,
12.0 and 11.6~MTU$^{-1}$. Damping has an especially simple form: since
$\overline{-2E_b}=-2L_b\bar q_b$, its contribution to the mean-logit
speed is exactly $-2/(1-\bar q_b)$, which depends only on the band's
saturation stage and grows without bound as $\bar q_b\to1$.

\begin{table}[t]
\caption{Time-mean contributions to $\dot\tau$ over the native ascent
intervals, obtained by trapezoidal integration divided by interval
length. Each column uses its own fitted rate vector. The complete
physical terms sum to the net clock speed; the variance correction of
the alternative reference-law decomposition is not added here.}
\label{tab:clockbudget}
\begin{tabular}{lrr}
\toprule
Contribution & Run U & Run C\\
\midrule
Truth-referenced production $\mathcal P$ & 2.173 & 6.168\\
Error--error redistribution $\mathcal N$ & 0.064 & $-0.021$\\
Interlevel exchange $\mathcal C$ & 0 & $-1.909$\\
Damping $-2E$ & $-1.214$ & $-3.146$\\
Net & 1.024 & 1.091\\
\bottomrule
\end{tabular}
\end{table}

The transverse projections show which physical balances bend the
band-growth trajectory. In Run C, the mean interlevel contributions to
$\dot\xi_1$ and $\dot\xi_2$ are $-1.41$ and $+0.27$ logit units per MTU:
the transformed exchange sink is strongest in the low band, which it
lowers relative to the others, and weaker in the middle band than in the
high band. The complete mean tendencies are, however, $+0.28$ and
$+0.13$: other terms can reinforce or outweigh an exchange contribution. The middle band
illustrates this distinction. Its time-mean transformed interlevel sink
is $-2.37$~MTU$^{-1}$, weaker than the low- and high-band values
$-5.12$ and $-3.36$. Its combined advective driving
$(\overline{\mathcal P}_b+\overline{\mathcal N}_b)/[L_b\bar q_b(1-\bar q_b)]$
is also smaller: 7.56 compared with 15.28 and 12.10~MTU$^{-1}$.
After damping, the mean logit velocities are 1.78, 2.52 and
2.00~MTU$^{-1}$ for the middle, low and high bands, respectively.
Relative to its own production, however, the middle band's exchange
sink is not weak: the ratio of transformed exchange to production is
$-0.35$, as in the low band ($-0.34$) and larger in magnitude than in
the high band ($-0.26$). Its slower advancement therefore reflects
weaker transformed driving, with an exchange sink in proportion to it,
rather than an anomalously strong sink. The error spectra (Fig.~\ref{fig:spectra}c,d) locate the persistent
logit offset at $k=6$--8, where the saturation spectrum peaks. There the
saturation ratio is about half its wavenumber average (0.56 of the mean
in the control and 0.50 in the fast-only experiment, in which the slow
level starts without error, compared with 0.93 in Run U). The deficit is
therefore maintained by the dynamics at the dominant wavenumbers rather
than inherited only from the analyses.

The alternative decomposition of Eq.~\eqref{eq:meanbudget} is verified
algebraically to within $7\times10^{-14}$ in logit tendency. Its
heterogeneity term averages $-4.17,-3.25,-3.47$~MTU$^{-1}$ in Run U and
$-2.06,-0.52,-1.06$~MTU$^{-1}$ in Run C, in band order. The consistently
defined remainder balances this reference-dependent correction. These
values quantify the averaging effect without treating it as an
additional loss term or identifying a Gompertz mechanism.

%% ======================================================================
\section{Discussion}\label{sec:discussion}
\subsection{Logit--Koopman coordinates within Koopman-based analysis}\label{sec:relation}
Linear inverse models (LIM), dynamic mode decomposition and their
Koopman generalizations describe geophysical variability through
observables whose evolution is linear.\cite{Penland1989,PenlandSardeshmukh1995,Navarra2024}
Their correspondence for matched sampling, centering and dictionaries is
established.\cite{Tu2014,Navarra2021} Logit--Koopman coordinates differ
from these analyses in three respects. The observable is the normalized error
energy of a chosen projection rather than a state anomaly, so the
coordinates describe predictability rather than variability. The
linearizing coordinate is prescribed by a reference growth law rather
than learned, and it has an explicit inverse to physical error energy.
And the coordinates are paired with an exact energy budget, so that
departures from the reference are attributed to physical terms instead
of being absorbed into a fitted operator. Correlations or
eigenfunction overlaps of observables, in either framework, measure
alignment; neither identifies a signed energy transfer without a budget.

The coordinates also extend the clock of linear modal growth through
saturation. For an autonomous linear system, a left-eigenvector
projection is a Koopman eigenfunction, and the energy of a single mode
grows at twice the real part of its exponent,\cite{Rowley2009,Brunton2022}
so that $\ln q$ is that mode's clock. Because $\eta=\ln q+q+O(q^2)$, the
logit coincides with this linear clock at small error, with $\lambda$ in
the role of twice the real part of the exponent, and continues it through
saturation, where no linear modal description applies. This continuation
is defined by the logistic reference; a modal decomposition alone does
not imply logistic saturation. Modes obtained from a LIM, dynamic mode
decomposition or Koopman analysis can nevertheless supply the
projections $\Pi_b$, provided band energies are evaluated in the
physical energy metric: such modes are in general nonorthogonal, so
$E_b$ then includes cross terms between modes, and the sum of squared
modal amplitudes in Eq.~\eqref{eq:modal} applies only to an
energy-orthonormal basis. The decomposition identifies which structures
carry error; the logit coordinates measure how the error energy in them
advances toward saturation and how their relative growth evolves.

\subsection{Growth clocks, reference laws and finite-size Lyapunov exponents}\label{sec:reflaw}
Equation~\eqref{eq:etaLambda} relates the logit speed to the
finite-amplitude growth rate. The clock also links lead time to error
amplitude, the variable of finite-size Lyapunov
analysis.\cite{Aurell1997,Boffetta2002} Under the reference law, an
error of amplitude $\delta=(qL)^{1/2}$ takes the time
$T_r=[\logit(r^2q)-\logit(q)]/\lambda$ to grow by a factor $r>1$, with
$r^2q<1$. The reference therefore implies the finite-size growth rate
\begin{equation}
 \lambda_\delta=\frac{\ln r}{T_r}
 =\frac{\lambda\ln r}{2\ln r+\ln\frac{1-q}{1-r^2q}} .
 \label{eq:fsle}
\end{equation}
It tends to $\lambda/2$, the amplitude growth rate of linear dynamics, as
$q\to0$ and vanishes as $r^2q\to1$. Departures of measured finite-size
exponents from Eq.~\eqref{eq:fsle} are therefore departures from the
reference clock, expressed in amplitude rather than lead time. The
present fixed-lead mean-curve rates do not implement threshold-passage
or rescaling protocols, and Eq.~\eqref{eq:meanbudget} quantifies one
difference between the clock of a mean error and those of individual
errors. The equal-error delays of Sec.~\ref{sec:exchange} are threshold
passages of launch-mean curves of total slow-level error, that is,
offsets of a scalar growth clock. They are not displacements purely
along the multiband longitudinal coordinate $\tau$, because the band
compositions of the two experiments differ at equal total error.

The amplitude dependence in Eq.~\eqref{eq:fsle} provides a
logistic-reference interpretation of the finite-amplitude filtering
demonstrated by Pe\~na and Kalnay.\cite{PenaKalnay2004} Choosing
perturbation size and rescaling interval from the system's physical
scales can reduce the influence of saturating fast components and reveal
slower coupled instabilities. In their fast-mode-dominated examples, the
leading Lyapunov and singular vectors did not achieve this separation.
Under the logistic reference, saturation reduces relative amplitude
growth while the logit clock advances uniformly; the finite-size rate in
Eq.~\eqref{eq:fsle} vanishes as the target amplitude approaches
saturation, $r^2q\to1$. Run C illustrates contrasting saturation stages,
with the fast levels initially near half their reference energies while
the slow level begins near 0.05 (Fig.~\ref{fig:spectra}). Whether
breeding isolates slow coupled modes in this regime remains to be tested.

The reference law is chosen a priori (Sec.~\ref{sec:theory}), and its
departures carry the physical information. Bach, Crisan and Ghil extend
the Dalcher--Kalnay law with multiplicative noise and note that a nonzero
fitted source can improve the growth-curve shape even in identical-twin
experiments without stochastic physics.\cite{BachCrisanGhil2025} A
finite-window fit of $\dot\eta=\alpha+s/q$ is thus an effective
description rather than an identification of fast-level injection, and
Eq.~\eqref{eq:meanbudget} shows that mean-clock departures can combine
physical tendencies with ensemble heterogeneity. In Logit--Koopman
coordinates these contributions are separated by the transformed budget
and tested by initial-error partitions, as in Sec.~\ref{sec:results}.

\subsection{Multiscale error propagation in clock coordinates}
The classical picture of multiscale predictability concerns upscale
influence: errors grow and saturate at fast, small scales and affect
slower, larger ones.\cite{Lorenz1969,RotunnoSnyder2008,DurranGingrich2014,SelzCraig2015}
The coordinates resolve two parts of that picture. The signed
interlevel budget measures net error-energy exchange along a trajectory;
the initial-error partitions test how fast-level perturbations influence
subsequent slow-level errors. A fast-to-slow seeding pathway coexists
with net slow-to-fast exchange in the control, even though the fast
levels are closer to saturation throughout the ascent
(Fig.~\ref{fig:spectra}). The two act on different attributes of
slow-level growth: the exchange is a negative contribution to the speed
at which the slow-level error advances, while fast-only initialization
yields a delayed slow-error growth curve, offsetting its scalar clock and
altering the band composition at equal total error. Linear-looking mean error growth and strong clock correlations
therefore need not identify the direction of net energy exchange. This
result concerns the archived initial errors of this Lorenz-96 regime and
does not rule out other upscale predictability mechanisms in geophysical
turbulence.

Within the $X$ ring, $\mathcal N_b$ isolates conservative error--error
redistribution in the truth-referenced budget, and truth--error
interactions in $\mathcal P_b$ also couple bands. The transformed sum of
these terms, exchange and dissipation determines both the longitudinal
speed and the transverse evolution. This complete balance gives the
geometry a physical interpretation that neither scalar curve shape nor
cross-band correlation supplies alone.

\subsection{Applications beyond the Lorenz-96 OSSEs}
The construction requires only a physically defined error energy for
each projection, a fixed saturation reference and a verifying truth or
analysis. In atmospheric applications the projections can be physical
modes as well as spectral bands. Normal-mode functions
distinguish rotational and inertia--gravity families with horizontal and
vertical structure,\cite{KasaharaPuri1981,Zagar2015MODES} and have been
used to estimate scale-dependent forecast-error growth.\cite{Zagar2017}
For an error expansion $\delta x=\sum_\nu a_\nu\Phi_\nu$ in an
energy-orthonormal basis, define
\begin{equation}
 E_b=\sum_{\nu\in b}|a_\nu|^2,\qquad
 q_b=E_b/L_b,\qquad\eta_b=\logit(q_b),
 \label{eq:modal}
\end{equation}
for groups $b$ of modes. The basis identifies which motions carry error;
the logit coordinates describe their normalized growth and relative
evolution; and the modal energy budget, transformed by
Eqs.~\eqref{eq:logitbudget}--\eqref{eq:meanbudget}, attributes the
clock and transverse motion to balanced and wave dynamics.

The same coordinates apply to the spectral bands used to study
scale-dependent predictability in convection-permitting and global
models,\cite{SelzCraig2015,Judt2018,Zhang2019} where clock offsets and
speeds would separate differences in initial error from differences in
growth, and to modes obtained from linear inverse or Koopman analyses of
observed variability.\cite{PenlandSardeshmukh1995,Navarra2024} A
probability interpretation offers a further route, to stochastic models
of individual error trajectories. Under a specified logistic-mobility
noise, the logit has constant drift and diffusion
(Appendix~\ref{sec:probabilityappendix}), which provides an analytic
reference for modeling departures in individual error energies and for
conditioning trajectory emulators.\cite{NathanielGentine2026} Each of
these applications requires its own references, populations and
budgets; the present mean-clock fits establish neither a Markov closure
nor a noise model.

\subsection{Limitations}
The empirical scope is two regimes of one idealized system, with fixed
normalizations and declared interior populations. The regimes differ in
coupling, climatology, observation error, assimilation performance and
initial error, so their contrast is descriptive rather than a
controlled intervention on the coupling; the Run C analyses are
underdispersive and leave the fast levels weakly constrained. The 200
forecasts per regime share one truth trajectory and overlap in time,
which limits the effective sample size. Block-bootstrap sensitivity
quantifies within-archive sampling variability, conditional on the fixed
climatological references and selected windows. The affine fits
are descriptive fits to mean trajectories, and nearly affine logits do
not identify logistic dynamics (Table~\ref{tab:transforms}). Band
energies discard phase and spatial position; errors that arise from the
displacement of coherent structures produce correlated band errors
through common spatial gradients and may be described more compactly
after symmetry reduction or shifted modal
decompositions.\cite{Budanur2015,Reiss2018} Finally, the logit is
undefined where individual errors exceed their references, which
restricts individual-forecast diagnostics to interior populations.

%% ======================================================================
\section{Conclusions}\label{sec:conclusions}
Logit--Koopman coordinates connect a scalar reference growth law to the
geometry of scale-resolved forecast errors. The logistic benchmark
provides exponential odds, affine logit clocks and an explicit physical
inverse. With several bands, a longitudinal coordinate measures shared
advancement and transverse coordinates retain departures from the
reference band-growth relation. Transforming the exact error-energy
budget into this frame identifies which physical terms advance the
clock and which bend the trajectory. A complementary reference-law
decomposition makes the averaging effect of ensemble heterogeneity
explicit.

The OSSEs demonstrate this construction with nearly affine mean logits,
distinct correlation structures and an evaluated physical budget. In
the coupled control, net interlevel error-energy exchange is from $X$
to $Y$ and from $Y$ to $Z$, and removes slow-level error energy at a
nearly steady relative rate. Fast-only initial perturbations nevertheless seed
slow-level errors immediately; their mean then approximately follows the
control growth curve with a nearly constant delay of about 0.7~MTU,
while its band composition at equal total error differs, so that similar
scalar histories conceal different spectral evolution. The middle band's slower
advancement reflects the complete physical balance, including weaker
transformed advective driving with a proportionate exchange sink and an
error deficit maintained at the dominant wavenumbers,
rather than a uniquely strong interlevel sink. These findings separate
growth timing, relative scale evolution and signed exchange within the
same trajectories. The analytical reference geometry supplies the
common coordinates in which those distinctions can be measured and
interpreted. Because the construction requires only a physically defined
error energy, a fixed saturation reference and a verifying state, the
same coordinates apply to modal, spectral or learned decompositions of
other forecast systems.

\begin{acknowledgments}
The author acknowledges useful discussion with Zoltan Toth, Joseph Tribbia, Nedjelka Zagar.
\end{acknowledgments}

\section*{Author declarations}
\subsection*{Conflict of interest}
The author has no conflicts to disclose.
\subsection*{Author contributions}
Malaquias Pe\~na: Conceptualization, methodology, software, validation,
formal analysis, investigation, data curation, visualization,
writing--original draft, and writing--review and editing.
\section*{Data availability}
The OSSE truth runs, analyses, forecast-error data, model and
assimilation code, and the analysis scripts that support the findings of
this study are available from the corresponding author upon reasonable
request.

\appendix
\section{Probability interpretation of the logit coordinate}
\label{sec:probabilityappendix}
The deterministic growth clock also admits a probability interpretation.
Suppose that the vector of individual normalized error energies follows
an It\^o diffusion on $(0,1)^d$,
\begin{equation}
 \dd\boldsymbol q=\boldsymbol b(\boldsymbol q,t)\,\dd t+
       \Sigma(\boldsymbol q,t)\,\dd\boldsymbol W_t,
 \qquad A=\Sigma\Sigma^{\mathsf T}.
 \label{app:closure}
\end{equation}
Equation~\eqref{app:closure} is a modeling assumption about individual
forecasts, not a result of this study. It is a Markov closure: the
future of $\boldsymbol q$ depends only on its present value, with the
influence of the unresolved state represented by the diffusion term. The
mean-clock fits of Sec.~\ref{sec:results} constrain neither the drift
$\boldsymbol b$ as a function of $\boldsymbol q$ nor the covariance $A$,
and they do not test the Markov property. Moreover, this appendix
concerns the logits of individual errors, $\logit(q_i)$, whereas
Sec.~\ref{sec:results} uses the logit of the launch mean,
$\logit(\mathbb E q_i)$; by Eq.~\eqref{eq:meanbudget} the two evolve
differently whenever forecasts differ. Assume sufficient regularity and
an interior solution, or stop the process at a specified domain exit.

For a smooth observable $f$, the backward generator and its forward
adjoint are
\begin{equation}
 \begin{aligned}
 \mathcal Lf&=\sum_i b_i\partial_i f+
                     \tfrac12\sum_{ij}A_{ij}\partial_i\partial_j f,\\
 \partial_t p&=\mathcal L^*p
   =-\sum_i\partial_i(b_ip)+
                     \tfrac12\sum_{ij}\partial_i\partial_j(A_{ij}p).
 \end{aligned}
 \label{app:duality}
\end{equation}
The stochastic Koopman evolution propagates conditional expectations of
observables and its adjoint propagates probability
densities.\cite{Klus2020Generator} For time-dependent coefficients this
is a two-parameter evolution family; the density equation presumes that
a density exists and suitable boundary or decay conditions hold. Without
diffusion it reduces to Liouville transport.

With $s_i=q_i(1-q_i)$ and $\eta_i=\logit(q_i)$, It\^o's formula gives
the exact interior transformation
\begin{equation}
 \begin{aligned}
 \dd\eta_i&=\widetilde b_i\,\dd t+
                       \sum_k\frac{\Sigma_{ik}}{s_i}\,\dd W_k,\\
 \widetilde b_i&=\frac{b_i}{s_i}+
                       \frac{(2q_i-1)A_{ii}}{2s_i^2},\qquad
 \widetilde A_{ij}=\frac{A_{ij}}{s_is_j},
 \end{aligned}
 \label{app:itotransform}
\end{equation}
with coefficients evaluated at $\boldsymbol q=\sigmoid(\boldsymbol\eta)$.
When densities exist, $p_\eta(\boldsymbol\eta,t)=p_q(\boldsymbol q,t)\prod_i s_i$.
Additive noise in $\boldsymbol q$ is amplified near the endpoints, so
the logit does not in general make both drift and diffusion constant.

%% Layout only: lets the reference list end on this page (review copy).
\enlargethispage{\baselineskip}
A compatible reference has logistic mobility and Stratonovich noise,
\begin{equation}
 \begin{aligned}
 \dd q_i&=s_i\Bigl(r_i\,\dd t+\sum_kB_{ik}\circ\dd W_k\Bigr),\\
 \dd\boldsymbol\eta&=\boldsymbol r\,\dd t+B\,\dd\boldsymbol W_t,
 \end{aligned}
 \label{app:mobilityreference}
\end{equation}
with constant $\boldsymbol r$ and $B$. This logistic-mobility
reference differs from multiplicative-noise forecast-error models whose
nominal saturation scale need not bound individual
trajectories.\cite{BachCrisanGhil2025} The Stratonovich chain rule gives
the second equation directly; the It\^o form in $q_i$ includes the drift
correction $\tfrac12(BB^{\mathsf T})_{ii}s_i(1-2q_i)$. Finite initial
logits therefore have Gaussian increments with mean $\boldsymbol r\,t$
and covariance $t\,BB^{\mathsf T}$, and their inverse logits remain
interior at all finite times. The zero-noise limit recovers the
deterministic clock. With noise, the conditional mean logit advances at
the reference rate while the conditional mean odds grow at rate
$r_i+(BB^{\mathsf T})_{ii}/2$. Noise spreads logits along the clock
and, unless its covariance lies along the reference direction, across
it, without implying a signed energy exchange. The inverse map has derivative at most 1/4, whereas the
forward logit derivative is $1/[q_i(1-q_i)]$; total variation and
relative entropy between probability laws are invariant under this
change of variables. A generative model conditioned on these clocks
would evolve in an auxiliary noise time distinct from the forecast lead
time, and reconstructing full fields would require the phases and
spatial structure discarded by band energies.

\bibliographystyle{aipnum4-2}
\bibliography{references}
\end{document}